\documentclass[%
reprint,
superscriptaddress,
 amsmath,amssymb,
 aps,
 prx,
floatfix,
]{revtex4-2}

\usepackage[dvipsnames,svgnames,x11names,hyperref]{xcolor}
\usepackage[english]{babel}
\usepackage{graphicx}
\usepackage{array}
\usepackage{amsmath,amsfonts,amssymb}

\newcommand{\Figref}[1]{Fig.~\ref{#1}}
\newcommand{\revision}[1]{\textcolor{black}{#1}}

\usepackage{xcolor} 

\usepackage{hyperref}
\usepackage{float}
\hypersetup{colorlinks, 
    linkcolor={blue!75!black!80!yellow},
    citecolor={blue!75!black!80!yellow}, 
    urlcolor={blue!75!black!80!yellow}
    }

\begin{document}
\title{Mapping Lamb, Stark and Purcell effects at a chromophore-picocavity junction with hyper-resolved fluorescence microscopy}

\author{Anna Ros\l{}awska}
\thanks{These two authors contributed equally\\anna.roslawska@ipcms.unistra.fr\\tomas.neuman@ipcms.unistra.fr}
\affiliation{Universit\'e de Strasbourg, CNRS, IPCMS, UMR 7504, F-67000 Strasbourg, France}
\author{Tom\'{a}\v{s} Neuman}
\thanks{These two authors contributed equally\\anna.roslawska@ipcms.unistra.fr\\tomas.neuman@ipcms.unistra.fr}
\affiliation{Universit\'e de Strasbourg, CNRS, IPCMS, UMR 7504, F-67000 Strasbourg, France}
\affiliation{Center for Materials Physics (CSIC-UPV/EHU) and DIPC, Paseo Manuel de Lardizabal 5, Donostia - San Sebasti\'{a}n 20018, Spain.}
\author{Benjamin Doppagne}
\affiliation{Universit\'e de Strasbourg, CNRS, IPCMS, UMR 7504, F-67000 Strasbourg, France}
\author{Andrei G.  Borisov}
\affiliation{Institut des Sciences Mol\'eculaires d'Orsay (ISMO), UMR 8214, CNRS, Universit\'e Paris-Saclay, 91405 Orsay Cedex, France.}
\author{Michelangelo Romeo}
\affiliation{Universit\'e de Strasbourg, CNRS, IPCMS, UMR 7504, F-67000 Strasbourg, France}
\author{Fabrice Scheurer}
\affiliation{Universit\'e de Strasbourg, CNRS, IPCMS, UMR 7504, F-67000 Strasbourg, France}
\author{Javier Aizpurua}
\affiliation{Center for Materials Physics (CSIC-UPV/EHU) and DIPC, Paseo Manuel de Lardizabal 5, Donostia - San Sebasti\'{a}n 20018, Spain.}
\author{Guillaume Schull}
\thanks{guillaume.schull@ipcms.unistra.fr}
\affiliation{Universit\'e de Strasbourg, CNRS, IPCMS, UMR 7504, F-67000 Strasbourg, France}

\begin{abstract}
The interactions between the excited states of a single chromophore with static and dynamic electric fields confined to a plasmonic cavity of picometer dimensions are investigated in a joint experimental and theoretical effort. In this configuration, the spatial extensions of the confined fields are smaller than the one of the molecular exciton, a property that is used to generate fluorescence maps of the chromophores with intra-molecular resolution. Theoretical simulations of the electrostatic and electrodynamic interactions occurring at the chromophore-picocavity junction are able to reproduce and interpret these hyper-resolved fluorescence maps, and reveal the key role played by subtle variations of Purcell, Lamb and Stark effects at the chromophore-picocavity junction.         
\end{abstract}

\date{\today}

\pacs{78.67.-n,78.60.Fi,68.37.Ef}

\maketitle

\section{Introduction}

Seminal experiments demonstrating the possibility to study optical properties of individual molecular chromophores \cite{Moerner1989,Orrit1990} gave rise to the development of super-resolution fluorescence techniques, revolutionizing the field of optical microscopy \cite{Huang2009,Hell2015}, and inspired the engineering of molecule-based quantum-optical devices \cite{Lounis2000, toninelli2020single}.
At the core of many of these applications stands the possibility to control the optical and electronic properties of molecules via static and dynamical electromagnetic interactions. 
In the weak-coupling limit, the presence of electromagnetic field fastens the radiative decay rate of the emitter (Purcell effect \cite{purcell1946spontaneous, novotny_hecht_2006}) and shifts its emission energy to the red (Lamb shift) \cite{Lamb1947,bethe1947,bethe1950,Power1966}. Electrostatic fields, on the other hand, cause Stark shifts \cite{stark1914, starkeffect2018} of the emission lines. These effects are nowadays well-understood in settings where the chromophore can be treated as a point-like emitter exposed to homogeneous external electromagnetic fields. 
The opposite limit, where the physical dimensions of the fields confinements are small or comparable with the size of the molecular emitter, remains largely unexplored.
Here, we address this extreme limit of light-matter interaction by analysing the spectra of light emission from a single electrically driven molecule in an atomically sharp tunneling junction.
We further show how these effects can be exploited to elucidate the electronic and optical properties of individual molecules and their interaction with a plasmonic cavity enabling ultimate spatial resolution by combining scanning tunneling microscopy (STM) and optical spectroscopy.  

In the respect of optical imaging of single molecules, scanning near-field optical microscopy \cite{Betzig1993, Richards2004ssnom, frey2004highresolution, Novotny2006, Kuhn2006enhancement, hartschuh2008tipenhanced, mauser2014tipenhanced, Park2018radiative} has proven to be a valuable tool thanks to its ability to confine electromagnetic fields at the extremity of the microscope tip.
This path has been explored to its limits in tip-enhanced Raman spectroscopy (TERS) \cite{Zhang2013,Lee2019} and tip-enhanced photoluminescence (TEPL) \cite{Yang2020subnanometer} to investigate vibronic and fluorescence signals with sub-molecular precision. Scanning tunneling microscope-induced luminescence (STML), which uses tunneling electron as an excitation source rather than photons, provided similar if not better, spatial resolution over fluorescence signals of chromophores \cite{Qiu2003,Chen2010,Chong2016, Chong2016ordinary,Zhang2016,Imada2016,Grosse2017,Doppagne2017,Imada2017,Zhang2017_vtomas,Zhang2017_a,Doppagne2018,Dolezal2019,Doppagne2020, Rai2020}. In typical STML experiments, hyper-resolved fluorescence microscopy (HRFM) maps are recorded by collecting the \textit{intensity} of the photon emission as a function of atomic-scale variations of the STM tip position with respect to the molecule. These HRFM maps, however, are challenging to interpret because the excitation and emission mechanisms driven by electronic and optical processes are simultaneously at play.

\revision{In contrast to these previous STML experiments, we show that HRFM maps of molecular electroluminescence lines embed information on the electromagnetic and electrostatic environment of the molecule encoded in the atomically-resolved line widths and line shifts.
Based on a comparison between experimental data and theoretical calculations obtained for a free-base phthalocyanine (H$_{2}$Pc) chromophore,} we demonstrate that line-width maps directly reflect the large Purcell effect generated by the coupling between the chromophore and gap plasmons at the tunneling junction, providing an access to variations of the molecular excited state lifetime with tip position. This signal therefore solely reflects the optical characteristics of the chromophore and can be used to generate sub-nanometric fluorescence maps. On the other hand, the energy of the chromophore fluorescence line depends on a subtle interplay between Lamb- \cite{Zhang2017_vtomas} and Stark-shifts \cite{Kuhnke2017a} induced by the presence of the tip. Whereas the former provides intimate details of the exciton-plasmon coupling, the latter reveals information about the transfer of charges associated with the chromophore electronic transition. In addition, our theoretical calculations highlight the decisive role played by the extreme confinement of the electromagnetic fields at the tip apex -- eventually acting as a cavity of atomic dimensions, or plasmonic picocavity \cite{Benz2016, Barbry2015, urbieta2018, Rossi2019, Yang2020subnanometer, wu2021, roslawska2021atomicscale} -- that is responsible for the ultimate spatial resolution in HRFM maps. The main advantage of our approach based on the mapping of the Purcell, Lamb, and Stark effects is that it is not hampered by the excitation probability of the chromophore, and solely reflects the optical (line-width maps) and electronic (line-shift map) characteristics of the exciton. 
Overall, this approach provides simultaneously electronic and optical signals of chromophores with close-to-atomic resolution, and decisive atomic-scale information on fundamental plasmon-exciton interactions in the limit where the spatial scales of the confined fields are comparable to that of the quantum emitter.

\noindent

\begin{figure*}
  \includegraphics[width=1.0\linewidth]{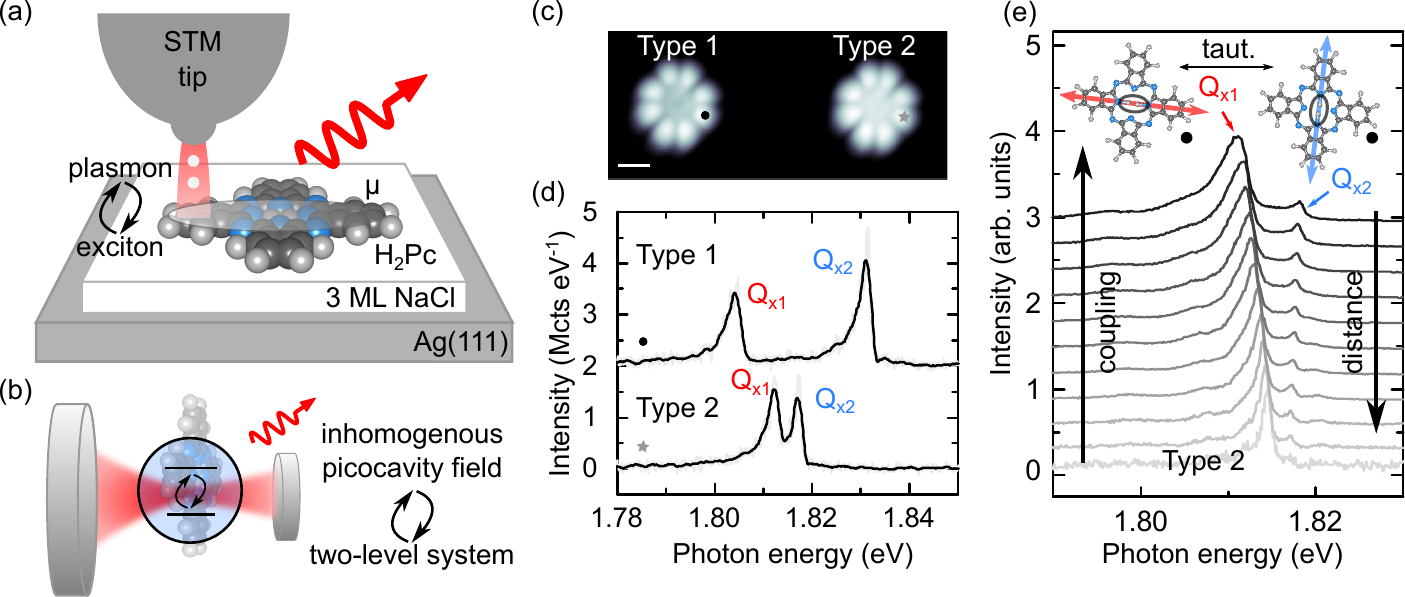}
  \caption{\label{fig1} (a) Sketch of the experiment where the plasmonic silver tip of an STM is used to excite the fluorescence of a single H$_{2}$Pc molecule deposited on a NaCl covered Ag(111) sample. The molecular dipole ($\mu$) is schematically marked by a grey ellipse. (b) Sketch of the system where an extended two-level system (molecule) interacts with the confined electromagnetic field of the plasmonic picocavity. (c) STM image ($V$ = -2.5 V, $I$ = 10 pA, scale bar: 1 nm) and (d) STML spectra ($V$ = -2.5 V, $I$ = 200 pA, $t$ = 120 s) of a type 1 and a type 2 H$_{2}$Pc (see main text for detail) acquired at positions marked by a black dot and grey star in (c). (e) Successive STML spectra acquired during the approach of the STM tip to a H$_{2}$Pc molecule of type 2. $I$ = 2 pA - 500 pA, $V$ = -2.5 V, $t$ = 30 s - 360 s. Inset: ball-and-stick models of two H$_{2}$Pc tautomers with the $Q_{x1}$ and $Q_{x2}$ dipoles marked by a red and blue arrow respectively. The hydrogen atoms are marked by ellipses. The black dot indicates the location of the tip for the experiment.}
\end{figure*} 

\begin{figure*}
  \includegraphics[width=.9\linewidth]{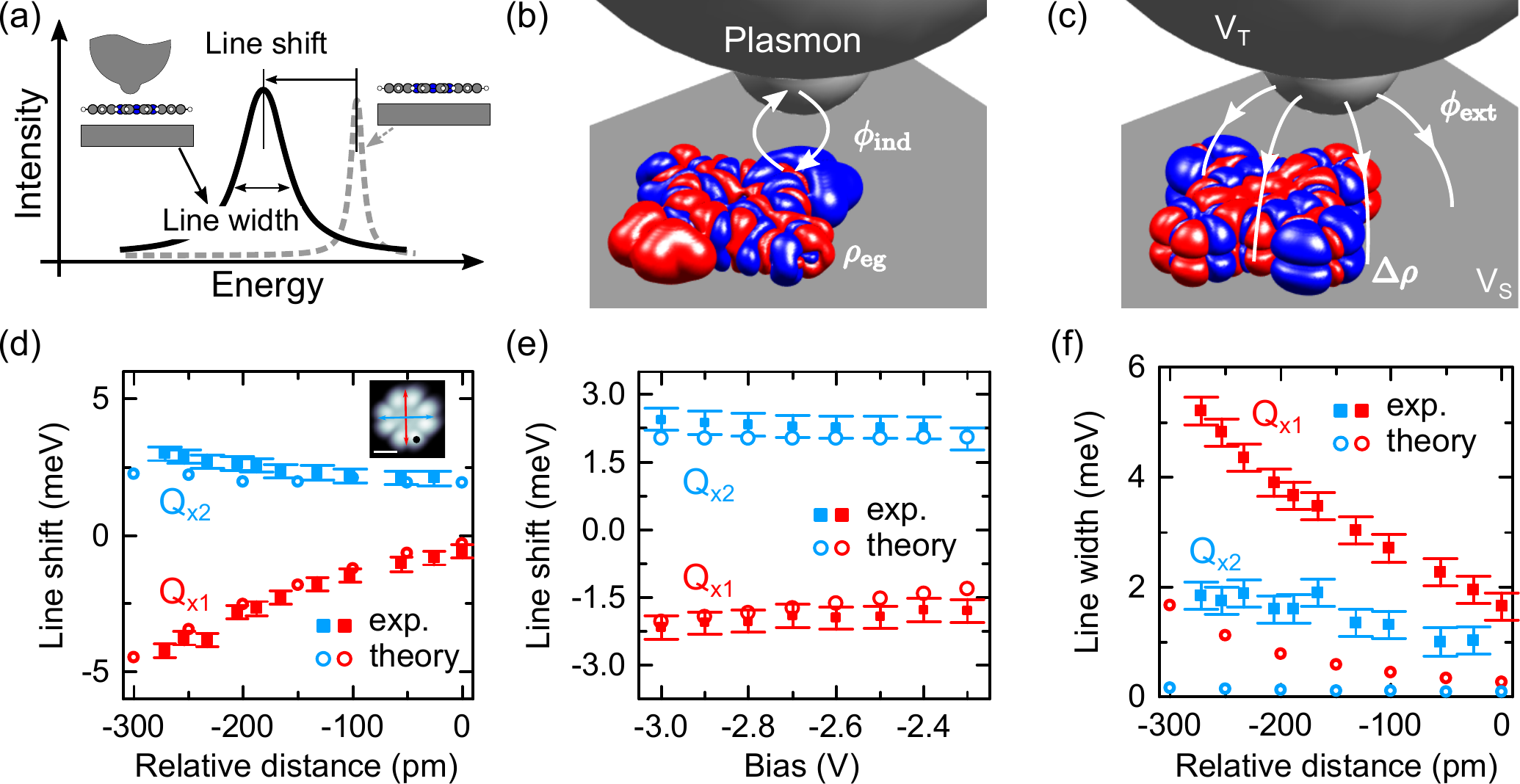}
  \caption{\label{fig2a} Probing and calculating the DC Stark effect, Lamb shift, and Purcell effect in STML. (a) Schematic representation of the overall shift and broadening of the excitonic photon emission spectral line recorded in STML as the tip approaches the sample molecule. The two schemes show the systems used to model the STM junction, the molecule is separated from the substrate by a vacuum gap of thickness d = 0.5 nm. (b) The excitonic transition charge density $\rho_{\rm eg}({\bf r})$ induces a plasmonic response in the tip and the substrate which is represented by the induced plasmonic potential $\phi_{\rm ind}$. (c) The difference between the electron densities of the molecular electronic excited state (Q$_x$), and the ground state, the difference charge density $\Delta\rho({\bf r})$, is exposed to the external potential $\phi_{\rm ext}({\bf r})$ generated by the voltage $V=V_{\rm T}-V_{\rm S}$ applied to the substrate, and the tip of the microscope. (d) Line shift as a function of the relative tip-sample distance ($V$ = -2.5 V). Inset: STM image $V$ = -2.5 V, $I$ = 10 pA, scale bar: 1 nm. The black dot marks the position of the measurements in (d-f). The color arrows indicate the two dipoles. (e) Line shift as a function of bias. The tip-sample distance is fixed during the measurement. The line shift in (d) and (e) was calculated taking photon energy $h\nu = 1.815$ eV as a reference value. (f) Line width as a function of the relative tip-sample distance ($V$ = -2.5 V).  The overall shift (Lamb shift combined with the Stark shift) (d, e) and width (f) of the excitonic emission line experimentally recorded (full squares) are compared with the theoretical calculations (empty circles) obtained using the approach presented in (a-c). The data in (d) and (f) are extracted from the measurement shown in \Figref{fig1}e.} 
\end{figure*} 

\section{Purcell, Lamb and Stark effects in STML}\label{sec:mol}

The experimental approach [\Figref{fig1}(a)] consists in using the tunneling electrons of an STM to excite the fluorescence of individual H$_{2}$Pc molecules separated from a Ag(111) surface by three insulating monolayers of NaCl. In this configuration, the insulating layer is sufficiently thick to prevent the quenching of the fluorescence due to the direct contact between chromophores and metallic surfaces. The sketch also represents the interaction between plasmons confined at the extremity of the STM tip and the transition dipole moment $\mu$ of the H$_{2}$Pc molecule. In this configuration, the electromagnetic field confined at the plasmonic picocavity created by the atomically sharp tip couples with a molecular two-level system of larger spatial dimensions [\Figref{fig1}(b)], in strong contrast to the homogeneous field distribution generally obtained in more standard electromagnetic cavities. 

The topographic STM image in \Figref{fig1}(c) shows two H$_{2}$Pc molecules adsorbed on a NaCl island. This image reveals a slightly dimmer contrast in the center of the left molecule (labelled type 1), as compared to the right molecule (labelled type 2). This behaviour has been attributed to slightly different geometries of the two molecules adsorbed on top of the NaCl ionic crystal \cite{Doppagne2020}. The difference in adsorption geometries also affects the fluorescence properties of the molecules as can be seen in the STML spectra of type 1 and 2 molecules provided in \Figref{fig1}(d). Here, the molecules were driven to their excited states by tunneling electrons at $V$ = -2.5 V applied to the sample and the tip-assisted light emission of these molecular excitons was detected in the far field. These high-resolution spectra show the main fluorescence transition, Q$_{x}$, of H$_{2}$Pc, which is associated with a dipole moment oriented along the two hydrogens of the central core of the molecule. In \Figref{fig1}(d), this transition appears as a doublet of lines labelled Q$_{x1}$ and Q$_{x2}$, whose energy separation is different for type 1 and type 2 molecules. This duplication of the Q$_{x}$ line has been attributed to the fast tautomerization of the central hydrogens within H$_{2}$Pc during the acquisition of an STML spectrum [\Figref{fig1}(e)] \cite{Doppagne2020}. The positions of hydrogen atoms are marked by ellipses in the inset of \Figref{fig1}(e). When recording time-integrated spectra, one simultaneously records the fluorescence of both tautomers. The STML spectra of type 1 molecules feature a large difference between the Q$_{x1}$ and Q$_{x2}$ excitonic lines which has been attributed to substantially different adsorption configurations of the two tautomers on the NaCl surface \cite{Doppagne2020}. Type 1 molecules can therefore be used to unambiguously separate the spectral contributions of Q$_{x1}$ and Q$_{x2}$, a property which we exploit in Section\,\ref{sec:map}.
As we detail in this paper (Section\,\ref{sec:deg}), the small energy difference in the case of type 2 molecules finds its origin in the coupling of the exciton with the electromagnetic fields induced in the presence of the STM tip.
The first evidence of the influence of this coupling is provided in \Figref{fig1}(e), where the spectral characteristics of a type 2 molecule are monitored as a function of the tip-molecule distance. 
By laterally positioning the tip as marked in \Figref{fig1}(e) we ensure that Q$_{x1}$ and Q$_{x2}$ experience non-equivalent interaction with the electromagnetic environment induced by the tip. 
Similarly to what was reported in \cite{Zhang2017_vtomas}, this distance-dependent spectrum reveals a shift of the Q$_{x1}$ line towards lower energies when the tip-molecule distance is reduced [\Figref{fig1}(e)]. In contrast, we observe a shift of the Q$_{x2}$ line towards higher energies. In both cases, these energy shifts are accompanied by a systematic enlargement of the Q$_{x1}$ and Q$_{x2}$ line widths.

To interpret these line shifts and line broadening we developed a theoretical approach accounting for the electromagnetic interactions occurring between the molecular chromophore and the plasmonic picocavity [\Figref{fig2a}(a)]. Additional details are provided in Appendix\,\ref{app:theory}.
First we consider the electron transition charge density $\rho_{\rm eg}$ for the optical transition at frequency $\omega_{\rm eg}$ between the ground (g) and excited (e) electronic states of the molecule, calculated using time-dependent density-functional theory (TD-DFT) \cite{valiev2010nwchem}. The transition charge density can be viewed as an oscillating electron charge acting as a source stimulating the plasmonic response of the cavity, which here takes the form of the induced plasmonic potential $\phi_{\rm ind}$ [\Figref{fig2a}(b)].
This potential then acts back on the exciton causing a line shift $\hbar\delta\omega_{\rm eg}={\rm Re}\lbrace\int\rho_{\rm eg}({\bf r})\phi_{\rm ind}({\bf r}){\rm d}^3{\bf r}\rbrace$ (Lamb shift), and broadens the excitonic emission line by $\hbar\gamma_{\rm eg}=-2{\rm Im}\lbrace\int\rho_{\rm eg}({\bf r})\phi_{\rm ind}({\bf r}){\rm d}^3{\bf r}\rbrace$ (Purcell effect). 
We additionally consider that the molecule is exposed to a strong inhomogeneous electrostatic potential $\phi_{\rm ext}$ resulting from the bias applied between the tip and the substrate [\Figref{fig2a}(c)]. This potential modifies the energies of the electronic ground and excited states via the DC Stark effect resulting in another shift of the excitonic energy $\delta E_{\rm St}=\int\Delta\rho\phi_{\rm ext}{\rm d}^3{\bf r}$, which is merely the energy difference resulting from the redistribution of charge (charge transfer) associated with the electronic excitation in the molecule. Here, $\Delta\rho=\rho_{\rm e}-\rho_{\rm g}$ is the difference charge density shown in \Figref{fig2a}(c), and $\rho_{\rm g}$ ($\rho_{\rm e}$) is the ground (excited) state charge density.

Armed with this theoretical framework, we compare the line shifts and line widths recorded experimentally 
(full squares) with the theoretical prediction (empty circles) and plot them as a function of the relative tip-sample distance in \Figref{fig2a}(d) and (f). 
These approach curves are plotted for two excitons Q$_{x1}$ and Q$_{x2}$, whose orientations with respect to the tip are non-equivalent, in the case of a type 2 H$_2$Pc molecule. We obtain the experimental data by fitting the spectra shown in \Figref{fig1}(e) using the procedure described in Appendix\,\ref{app:fits}.

As the experimentally available tip-sample distance and line shifts are relative, in the theoretically calculated data we choose the origin of the tip-substrate distance (i.e. the gap size) guided by the measured values and infer from the experimental data that the frequency of the exciton of the molecule on the substrate unperturbed by the tip is $\sim 1.815$\,eV as detailed in Appendix\,\ref{app:StarkLamb}. 
We observe two qualitatively different trends for the total line shift shown in \Figref{fig2a}(d) for Q$_{x1}$ (red markers) and Q$_{x2}$ (blue markers) excitons as a function of the relative tip-sample distance. 
The excitonic line of Q$_{x1}$ is red shifting as the tip approaches while the opposite trend is observed for Q$_{x2}$. 
As we detail in Appendix\,\ref{app:StarkLamb}, in our setup the Lamb shift always redshifts the exciton frequency irrespective of the tip position. The Stark shift can either red- or blue-shift the exciton frequency depending on the STM tip position. In Appendix\,\ref{app:StarkLamb} we show that the Stark and Lamb shift jointly lower the frequency of Q$_{x1}$ for the tip position marked by a black dot in the inset of \Figref{fig2a}(d). For the same tip position, the Stark effect blue shifts the Q$_{x2}$ exciton and overrides the Lamb effect.
The role played by the Stark shift is confirmed experimentally [see \Figref{fig2a}(e) and Appendix\,\ref{app:starkexperiment}] by monitoring the Q$_{x1}$ and Q$_{x2}$ line positions as a function of voltage while keeping the tip-molecule distance fixed. In this configuration, the plasmon-exciton coupling is constant. The line shifts in that situation are therefore exclusively related to the increasing electrical field, and thus directly measure the Stark shift. This effect is accounted for by our model [empty circles in \Figref{fig2a}(e)].

In \Figref{fig2a}(f), experiment and theory also report an increasing line width with decreasing tip-sample distance for Q$_{x1}$ and Q$_{x2}$. In the theory, this reflects the increased decay rate of the exciton into plasmons by the Purcell effect. The experiments reveal a similar behaviour, with an offset of $\approx$1 meV suggesting an additional decoherence channel not accounted for in the simulations (e.g., interaction of the excitons with substrate phonons or low-energy vibrations). As the experimental tendencies are well reproduced by our theoretical predictions, we can tentatively assign the component of the line broadening that depends on the tip position to the plasmonic Purcell effect. Overall, this indicates an excited state lifetime of the order of 1\,ps, about three orders of magnitude faster than what is expected for the free molecule \cite{SIBATA2004131, Caplins2016},
and highlights the strong amplification of the chromophore emission provided by the plasmonic picocavity.    
The plasmon-induced line broadening recorded for Q$_{x1}$ is considerably larger than for Q$_{x2}$ as the relative orientations of the excitonic dipolar moments with respect to the tip differ for the two excitons, resulting in larger plasmon-exciton coupling strengths for Q$_{x1}$ than for Q$_{x2}$. 
Importantly, the plasmon-induced broadening of the excitonic line is expected to be free of the influence of the electrostatic field (which only results in the energy shift of the line). The modification of the line width therefore provides a direct measure of the plasmon-exciton coupling.  

\begin{figure*}
  \includegraphics[width=0.7\linewidth]{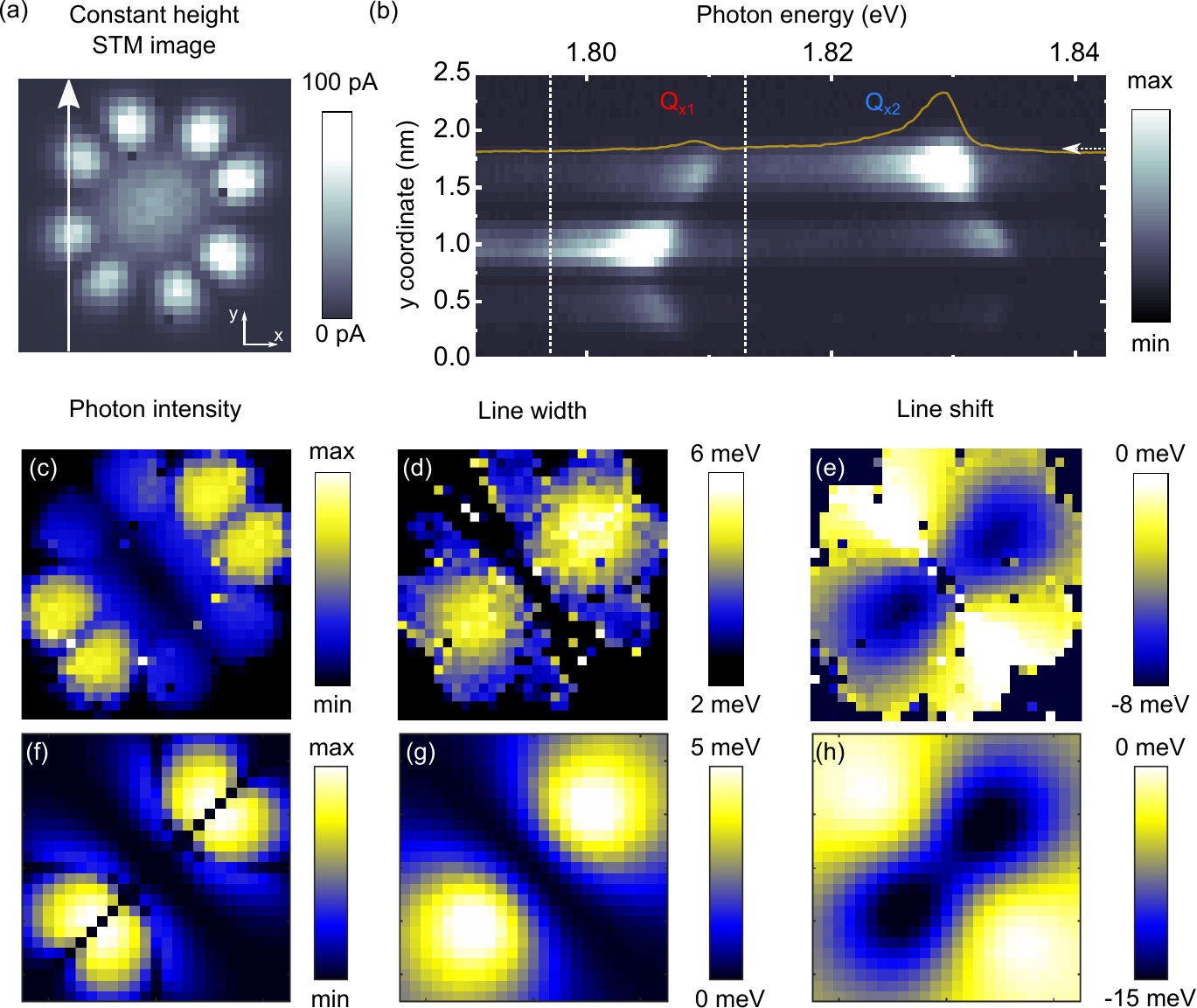}
  \caption{\label{fig2} HRFM mapping of a type 1 H$_{2}$Pc molecule. (a) Constant height STM current image ($V$ = -2.5 V, 2.5 $\times$ 2.5 nm$^2$, $I_{max}$ = 101 pA) of a type 1 H$_{2}$Pc molecule acquired with an open feedback loop (constant height) \cite{Doppagne2020}. (b) 2D representation of successive STML spectra acquired for STM tip positions along the white arrow in (a). STML spectrum corresponding to the dashed white arrow is plotted as a yellow curve. (c) Photon intensity \cite{Doppagne2020}, (d) line-width and (e) line-shift maps of the Q$_{x1}$ exciton reconstructed from STML spectra acquired for each pixel of the map in (a). The integration range (1.797 - 1.813 eV) is indicated by white dashed lines in (b). Theory maps of (f) the photon intensity, (g) line width, and (h) line shift assuming a 1\,nm gap between the tip and the substrate and considering a picocavity tip. We present the theory maps consistently with the experiment and measure the line shift relatively to the maximal value of the exciton energy (the line shift thus appears to be negative). The molecule is positioned $0.5$\,nm above the silver substrate (see the configuration in \Figref{sfig:geometry} in Appendix\,\ref{app:theory}).} 
\end{figure*} 

\begin{figure}
  \includegraphics[width=1\linewidth]{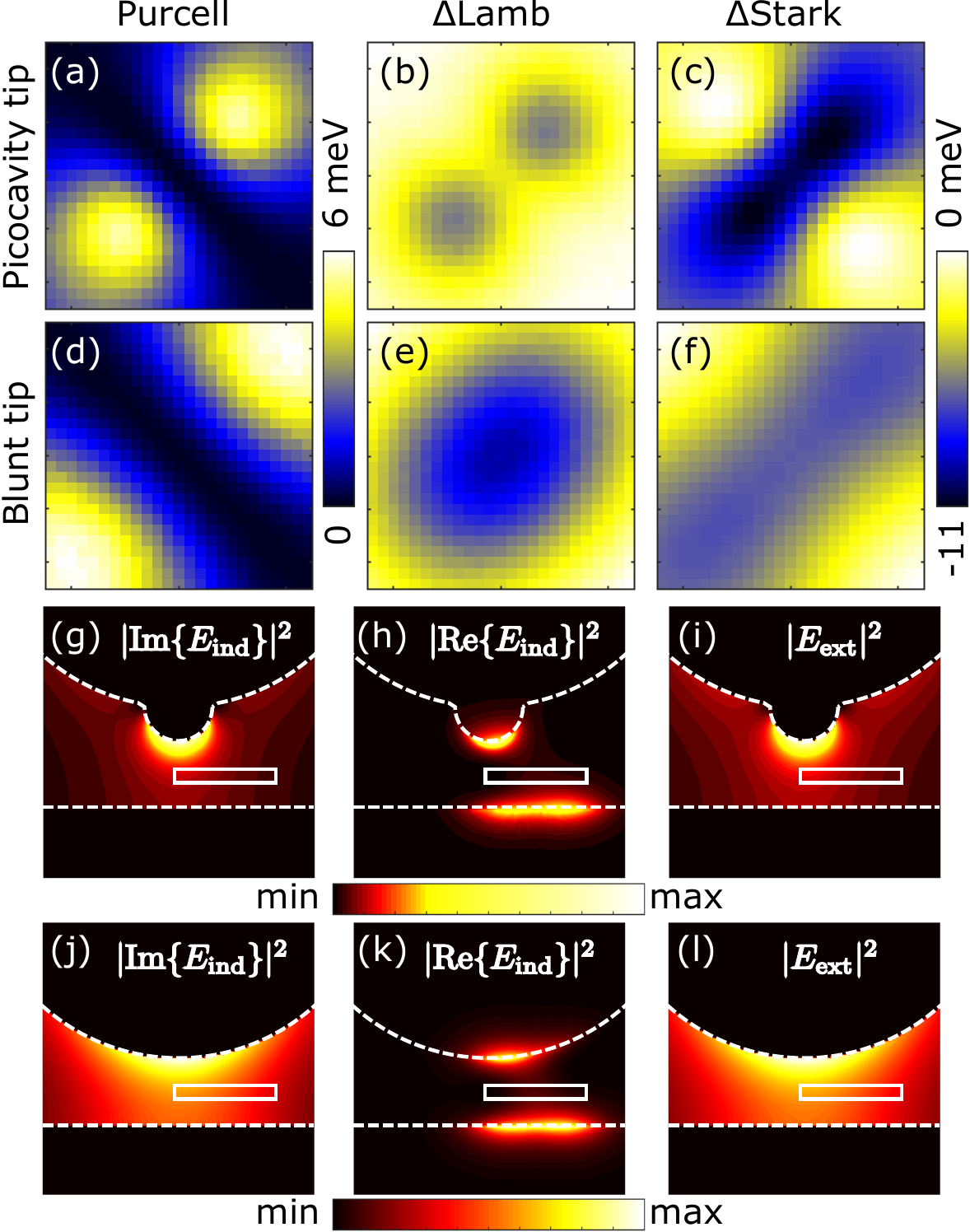}
  \caption{\label{fig3_theory} Dissecting the line broadening and line shift of a single-molecule exciton in an STM plasmonic cavity. Maps of (a,d) the line width, and the variation of (b,e) the Lamb shift and (c,f) the Stark shift induced in (a-c) plasmonic picocavity, and (d-f) cavity formed between blunt tip and the substrate. The values of the shifts are displayed in relative units with respect to the maximal value obtained in the maps. The size of the mapped area is $2.5\times2.5$ nm$^2$. The color scales in (a,d), and (b,c,e,f), respectively, are identical. (g,j) $|{\rm Im}\lbrace {\bf E}_{\rm ind}\rbrace |^2$ and (h,k) $|{\rm Re}\lbrace {\bf E}_{\rm ind}\rbrace |^2$ with ${\bf E}_{\rm ind}=-\nabla\phi_{\rm ind}$. The induced fields are generated by the Q$_x$ exciton of a molecule whose position is marked by the white dashed rectangle. (i,l) shows $|{\bf E}_{\rm exc}|^2=|\nabla\phi_{\rm ext}|^2$. Calculations in (g,h,i), and (j,k,l) have been done for the picocavity tip, and the blunt tip, respectively. All the fields are displayed in the plane $y=0$ intersecting the tip axis. The size of the displayed region is $4\times4$ nm$^2$.} 
\end{figure} 
 
\section{Mapping the Purcell, Lamb and Stark effect}\label{sec:map}

The shifts and broadening of the fluorescence lines induced by the Stark, Lamb, and Purcell effects suggest a modality of STML based on the mapping of these signals. This technique could be used to acquire hyper-resolved information on the molecular excitons and their coupling to the picocavity plasmons.

To demonstrate this capability, we have recorded fluorescence spectra at each pixel of a constant height STM current image [\Figref{fig2}(a)] of a type 1 H$_{2}$Pc. As only the effect of the lateral position of the tip is of interest here, the map was recorded in constant height mode to avoid any changes of the tip-molecule distance that would affect the measurement of the line shift and line width. 
 
\Figref{fig2}(b) is a 2D representation of successive spectra acquired along the vertical arrow in \Figref{fig2}(a) where the Q$_{x1}$ and Q$_{x2}$ contributions can be identified. Similarly to what was reported in \Figref{fig1}(e), the intensity, energy position and width of the fluorescence lines vary as a function of the lateral position of the tip with respect to the molecule. As the energy separation between the Q$_{x1}$ and Q$_{x2}$ lines of type 1 molecules is substantially larger than their line shifts, the analysis of one peak is not perturbed by the presence of the other. In \Figref{fig2}(c) to (e) we show the position-dependent maps of the Q$_{x1}$ line characteristics recorded simultaneously with the current map in \Figref{fig2}(a). The spectral range of the emission is marked by the white dashed lines in \Figref{fig2}(b). In \Figref{fig2}(c) one first identifies lobes that are characteristic of the highest occupied molecular orbital (HOMO) of the molecule. 
One also notices a dark node that crosses the molecule from the top left to the bottom right corner of the \Figref{fig2}(c) photon intensity map.  
In \Figref{fig2}(d) and (e) we display the evolution of the line width and the energy of the Q$_{x1}$ exciton, respectively, for each position of the tip with respect to the molecule. These images reveal similar two-lobe patterns oriented along the Q$_{x1}$ transition dipole moments of the molecule where the width of the emission line is broader and the line red-shifted compared to spectra recorded aside from the lobes. However, while these lobes appear rather homogeneous and round in the line-width map, they adopt a more elongated shape in the line-shift one. Similar data, tilted by 90$^\circ$, are recorded for the  Q$_{x2}$ contribution (see Appendix\,\ref{app:qx2}).    

To interpret these results, we apply our theoretical model considering the picocavity formed between the silver substrate and an atomic scale protrusion on the silver STM tip (see Appendix\,\ref{app:theory}) and calculate constant-height maps of the light intensity emitted by the electrically pumped molecule [\Figref{fig2}(f)], plasmon-induced line width [\Figref{fig2}(g)], and line shift [\Figref{fig2}(h)]. The theoretically calculated maps are in excellent agreement with the experiment. 
As we discussed in \cite{Doppagne2020}, in the intensity maps [\Figref{fig2}(c,f)] the information about the plasmon-exciton interaction is inherently convoluted with experimental features arising from the electron tunneling, which are proportional to the density of states associated with the HOMO of the H$_{2}$Pc molecule. 
\revision{Our theoretical Purcell map reproduces both the pattern and the amplitude of the changes of our experimental line-width maps. Based on this agreement,} we attribute the line-width maps in \Figref{fig2}(d,g), to the changes of the chromophore's excited state lifetime as a function of tip position. The line-width maps can therefore be interpreted as an artefact-free optical image of the chromophore. 
Finally, the maps of the relative line shifts shown in \Figref{fig2}(e,h) reveal an elongated two-lobe pattern featuring the same symmetry as the one obtained in \Figref{fig2}(d,g). However, in this map, the electrostatic and electrodynamic contributions to the exciton energy shift are entangled. 

To further elucidate the physical effects underlying experimental and theoretical maps in \Figref{fig2}, we analyze the respective role of the plasmonic Purcell effect, the Lamb shift, and the Stark effect for both a sharp picocavity tip [\Figref{fig3_theory}(a-c)], and a blunt tip missing the sharp protrusion [\Figref{fig3_theory}(d-f)]. The line broadening due to the Purcell effect [\Figref{fig3_theory}(a,d)] features the two-lobe structure for both tip shapes but is considerably delocalised when the blunt tip is considered. This points towards the importance of the picocavity in achieving HRFM. In [\Figref{fig3_theory}(b,e)] and  [\Figref{fig3_theory}(c,f)] we show the Lamb and Stark contributions to the relative line-shift maps, respectively. For the picocavity tip, the Lamb shift map features a two-lobe pattern whereas the Stark shift map reports a more elongated structure that strongly resembles the experimental data. These data also reveal the smaller magnitude of the Lamb shift variations and indicates the prevalence of the Stark effect in the line-shift map. 
On the other hand, when the blunt tip is considered, the Lamb and Stark shift variations are comparable in magnitude. As for the Purcell effect, their maps revealed considerably less resolved patterns than for the picocavity tip, where the two-lobe structures cannot be recognized anymore.    

Despite both being the result of the plasmon-exciton interaction, the patterns observed in the maps of the Lamb shift are qualitatively different from the maps of the line width. This is especially apparent for the blunt tip. The origin of this effect can be understood from the spatial distribution of the imaginary part of the induced plasmonic field in the cavity ($|{\rm Im}\lbrace {\bf E}_{\rm ind}\rbrace|^2$, with ${\bf E}_{\rm ind}=-\nabla\phi_{\rm ind}$) responsible for the line broadening [\Figref{fig3_theory}(g,j)], and its real part ($|{\rm Re}\lbrace{\bf E}_{\rm ind}\rbrace|^2$) responsible for the Lamb shift [\Figref{fig3_theory}(h,k)]. 
We consider the picocavity tip in \Figref{fig3_theory}(g-i), and the blunt tip in \Figref{fig3_theory}(j-l). 
The spatial distribution of the imaginary part of the induced plasmonic field corresponds to the field intensity generated by the dipolar plasmonic mode oscillating along $z$, which we tune to be resonant with the exciton by adjusting the tip geometry as detailed in Appendix\,\ref{app:theory}. This field distribution is an intrinsic property of the plasmonic picocavity and is therefore independent of the position of the source molecule. This single resonant dipolar plasmonic mode thus dominantly contributes to the Purcell effect and in turn mediates the far-field excitonic radiation recorded in STML. 
On the contrary, the real part of the induced field is strongly concentrated on the metal surfaces closely surrounding the molecule (see the bright spot on the metal surfaces below and above the molecule) and depends strongly on the molecular position (marked by the white rectangle). This localization is caused by the blue-detuned higher-order surface plasmon modes acting collectively as a plasmonic pseudomode that red-shifts the excitonic peak \cite{Delga_2014, Neuman2018}. 
The role of the picocavity, in this case, is to further localize the plasmonic response to the protrusion which is thus sensitive to the spatial variation of the excitonic transition-charge density, highlighting that the chromophore cannot be considered as a simple point dipole \cite{Neuman2018, Rossi2019}. 
Finally, $\phi_{\rm ext}$ features a strong gradient (electric field) ${\bf E}_{\rm ext}=-\nabla\phi_{\rm ext}$ in the $z$ direction due to the bias voltage. We show the magnitude of the electric field in \Figref{fig3_theory}(i) for the picocavity tip and \Figref{fig3_theory}(l) for the blunt tip. The protrusion of the picocavity tip additionally generates a strong gradient of $\phi_{\rm ext}$ in the plane of the substrate which is significant even across the geometrical extent of the molecule ($\sim1.5$ nm). It is this spatial inhomogeneity that results in the strong dependence of the DC Stark shift on the relative position of the molecule with respect to the tip. 

\begin{figure}
  \includegraphics[width=1\linewidth]{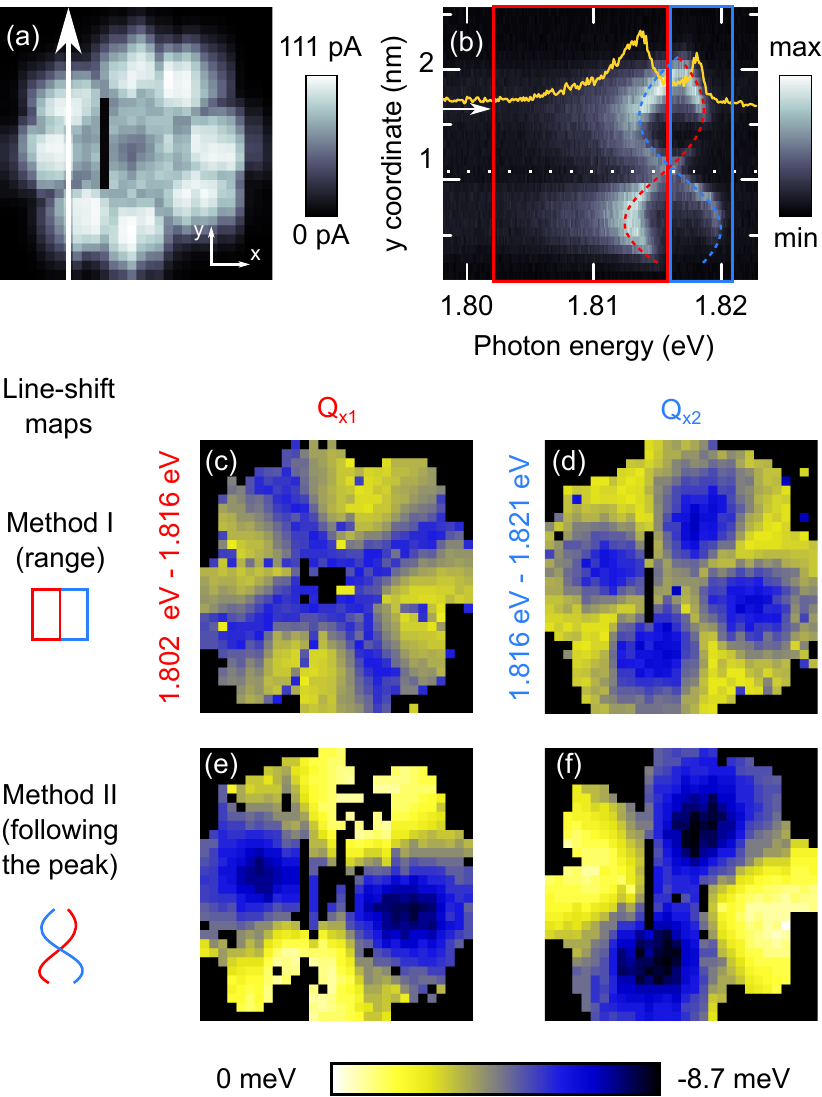}
  \caption{\label{fig3} HRFM mapping of a type 2 H$_{2}$Pc molecule. (a) Constant height STM current image ($V$ = -2.5 V, 2.6 $\times$ 2.6 nm$^2$, $I_{max}$ = 111 pA) of a type 2 H$_{2}$Pc molecule acquired with an open feedback loop. (b) 2D representation of successive STML spectra acquired for STM tip positions along the white arrow in (a). A spectrum corresponding to the position marked by a short white arrow is plotted in yellow. (c-f) HRFM line-shift maps of Q$_{x1}$ (c,e) and  Q$_{x2}$ (d,f) acquired for each pixel of the map in (a). The maps in (c,d) are obtained by selecting a range defined by the colored boxes in (b). The maps in (e,f) are obtained by following the given Q$_{x}$ exciton (dashed colored lines). Note the common color scale for all line-shift maps.} 
\end{figure} 

\section{HRFM of degenerate excitons}\label{sec:deg}

Having fully addressed the nature of the contrasts in HRFM maps of a type 1 molecule, we eventually apply this approach to a H$_{2}$Pc molecule of type 2 [\Figref{fig3}(a)] that displays a smaller energy separation between the Q$_{x1}$ and Q$_{x2}$ lines. \Figref{fig3}(b) is the 2D representation of successive spectra acquired along the vertical arrow in \Figref{fig3}(a), analogue to that in \Figref{fig2}(b) for the type 1 molecule. The yellow curve overlaid on the plot represents the STML spectrum indicated by a short white arrow where Q$_{x1}$ and Q$_{x2}$ contributions are identified. Similarly to what is reported for the type 1 molecule, the 2D plot in \Figref{fig3}(b) reveals an energy shift of the Q$_{x1}$ and Q$_{x2}$ lines as a function of the tip position. However, as the energy separation between the Q$_{x1}$ and Q$_{x2}$ lines is smaller for type 2 molecule, a configuration is reached (white dotted line) where the two contributions are degenerate. For larger and smaller values of $y$, the two spectral contributions can be separated again. Following an approach similar to that used in \Figref{fig2}, in \Figref{fig3}(c) and (d) we generated two line-shift maps by estimating the energy of the intensity maxima in the red and blue energy windows overlaid on \Figref{fig3}(b). In this case, one assumes that the energy of the Q$_{x1}$ line (Q$_{x2}$ line) in \Figref{fig3}(b) is always smaller (higher) than 1.816 eV, the energy at which the two peaks appear degenerate. In contrast with the maps obtained on type 1 H$_2$Pc, the line-shift maps reveal two different patterns of an unexpected four-fold symmetry. We also generated line-shift maps assuming that the Q$_{x1}$ line (Q$_{x2}$ line) keeps shifting to higher (lower) energy after crossing the position corresponding to degenerate Q$_{x1}$ and Q$_{x2}$ contributions [marked by a dashed line in \Figref{fig3}(b)]. This means considering that the energy of the Q$_{x1}$ line (Q$_{x2}$ line) follows the red (blue) dashed curve. Line-shifts maps generated with this method reveal the characteristic two-fold symmetry patterns, tilted 90$^{\circ}$ from each other, expected for the Q$_{x1}$ and Q$_{x2}$ contributions. From this analysis, we conclude that type 2 molecules naturally exhibit degenerate Q$_{x1}$ and Q$_{x2}$ transitions, a degeneracy that is lifted by the couplings of the tip with each of the two excited states whose strengths vary as a function of the lateral tip position. 

\section{Conclusions and outlook}

In this manuscript, we decipher the role played by electrostatic and electromagnetic interactions contributing to the unmatched spatial resolution obtained in STM-induced luminescence measurements of single molecules. This enables realizing hyper-resolved fluorescence mapping of two quantities: the width and the shift of fluorescence lines that, in contrast to more standard intensity maps, provide information completely free of features related to the electron excitation efficiency. 
These line-shift and width maps unravel the interaction between the excited states of a chromophore with the electrodynamic fields confined at the plasmonic picocavity, revealing the emergence of the Purcell, Lamb, and Stark effects. Their respective impact on the fluorescence maps is dissected with an approach that combines ab-initio and electrodynamical calculations. 

We foresee that \revision{these approaches may be used to reveal new characteristics of molecular systems studied in former STML works discussing line intensity mapping, in particular those where emission lines are very close in energy. They may also} open a venue to understand and design correlated studies of optical and electrical properties mediated by a single molecule in a junction. In addition to the information provided by standard STM, the HRFM photon intensity maps reflect the microscopic mechanism of electrical exciton pumping \cite{Doppagne2020}, the maps of line widths reveal optical properties of the exciton, and the Stark shift maps probe the redistribution of the electron charge between the ground and the excited state and can be well adapted to the study of molecular charge-transfer excitons. \revision{This strong Stark effect can be further employed to either gate the energy of individual excitons, for example in coupled multi-molecular architectures \cite{Zhang2016,Imada2016,Cao2021}, or sense local electrostatic fields. Similarly, sensing of the electromagnetic environment can be obtained by monitoring the Purcell effect. Microscopically controlled Stark shift of excitons has also been proposed to be at the core of molecular optomechanical devices \cite{Dutreix2020}, quantum transducers between superconducting qubits and optical photons \cite{Sumanta2017} or demonstrated to control strong coupling \cite{Laucht2009}.} The tool set provided by the modalities of STML demonstrated in this \textit{Article} is therefore particularly suited to study the basic building blocks of next-generation organic light-emitting diodes, organic solar cells, or even \revision{molecular} sources of non-classical light \revision{and molecular quantum technologies in general \cite{toninelli2020single}}.

%\\

\noindent
\section{Methods}

The STM data were acquired with a low temperature (4.5\,K) ultrahigh-vacuum Omicron setup adapted to detect the light emitted at the tip-sample junction. The optical detection setup is composed of a spectrograph coupled to a CCD camera and provides a spectral resolution of $\approx$ 0.2 nm \cite{Doppagne2020}. Tungsten STM-tips were introduced in the sample to cover them with silver to tune their plasmonic response. The Ag(111) substrates were cleaned with successive sputtering and annealing cycles. Approximately 0.5 monolayer of NaCl was sublimed on Ag(111) kept at room temperature, forming square bi- and tri-layers after mild post-annealing. Free-base Phthalocyanine (H$_{2}$Pc) molecules were evaporated on the cold ($\approx$ 5 K) NaCl/Ag(111) sample in the STM chamber. As the acquisition of each optical spectrum of the HRFM maps requires a long accumulation time, an atomic tracking procedure is used to correct the x,y,z position of the STM tip between the acquisition of each pixel, ensuring a full correction of residual thermal drifts that may have detrimental effects on STM measurements requiring long acquisition time.

\section*{Acknowledgements}
The authors thank Virginie Speisser for technical support and Alex Boeglin and Hervé Bulou for discussions. This project has received funding from the European Research Council (ERC) under the European Union's Horizon 2020 research and innovation program (grant agreement No 771850) and the European Union's Horizon 2020 research and innovation program under the Marie Sk\l odowska-Curie grant agreement No 894434. The Labex NIE (Contract No. ANR-11-LABX-0058\_NIE), the International Center for Frontier Research in Chemistry (FRC), the Spanish Ministry of Science (Project No PID2019-107432GB-I00) and the Department of Education of the Basque Government (Project No IT1164-19) are acknowledged for financial support.

\section*{Appendices}
\appendix

\section{Theoretical modelling of the Lamb, Stark, and Purcell effect in a plasmonic picocavity}\label{app:theory}

\noindent
The electronic structure of a free-base phthalocyanine is approximated as an electronic two-level system composed of electronic ground $|{\rm g}\rangle$ and excited $|{\rm e}\rangle$ states. The excited electronic state in this case represents the Q$_x$ electronic transition of the molecule. We assume that the 90$^\circ$ tilted exciton of the molecule, Q$_y$, is sufficiently energetically separated from the Q$_x$ excitonic transition so that neither the plasmonic interactions nor the Stark effect cause significant mixing of the two. The molecular excited and ground electronic states interact (i) with the static potential $\phi_{\rm ext}({\bf r})$ due to the bias voltage applied between the tip and the substrate, and (ii) with the dynamical plasmonic potential $\phi_{\rm ind}$ induced at the junction between the tip and substrate electrodes via their induced charge density. The electron charge density is represented by the operator $\hat{\rho}({\bf r})$, which can be expressed in the second quantization as:
\begin{align}
    \hat{\rho}({\bf r})=\sum_{ij, \alpha}\rho_{ ij}({\bf r})\hat{c}^\dagger_{i\alpha}\hat{c}_{j\alpha}\label{eq:densityoperator}
\end{align}
where $\hat{c}_{i\alpha}$ ($\hat{c}^\dagger_{i\alpha}$) are the fermionic single-particle annihilation (creation) operators with the indexes $i,\,j$ running over spatial orbitals and $\alpha$ runs over the spin indices $\uparrow$ and $\downarrow$. The densities $\rho_{ij}({\bf r})$ are defined as $\rho_{ij}({\bf r})=-|{\rm e}|\psi_i({\bf r})^\ast\psi_j({\bf r})$ with $\psi_i({\bf r})$ being the respective single-electron spatial orbitals, and ${\rm e}$ the electron charge.

To calculate the molecular ground and excited states we use the linear-response time-dependent density functional theory using the B3LYP functional and 6-31g* basis set, as implemented in NWChem \cite{valiev2010nwchem}. We obtain the excitations in the Tamm-Dancoff (configuration interaction singles -- CIS) approximation. 
In this approximation the excited states are defined as linear combinations of electron-hole pairs created on top of the filled Fermi sea defined by the ground state:
\begin{align}
    |{\rm g}\rangle = \prod_{a\in{\rm occup.}}\hat{c}^\dagger_{a\uparrow}\prod_{a\in{\rm occup.}}\hat{c}^\dagger_{a\downarrow}|{0}\rangle,
\end{align}
where $|0\rangle$ is the empty vacuum state. Further on we use the convention that $a$, $b$,... run over the occupied states of the ground state, and $i$, $j$, ... run over the unoccupied states. The excited state $|{\rm e}\rangle$ is generated from the ground state in the Tamm-Dancoff approximation as:
\begin{align}
    |{\rm e}\rangle=\sum_{ia\alpha}C_{ia,\alpha}\hat{c}^\dagger_{i\alpha} \hat{c}_{a\alpha}|{\rm g}\rangle.\label{eq:excited}
\end{align}
Here $C_{ia,\alpha}$ are coefficients fulfilling $\sum_{ia\alpha}\vert C_{ia,\alpha}\vert^2=1$, and for singlet excitations we have $C_{ia,\downarrow}=C_{ia,\uparrow}=C_{ia}$.

\subsection{The DC Stark effect in plasmonic picocavities}

The DC Stark effect commonly refers to the shift of the energies of quantum states of a system exposed to an external static electric field. The interaction energy of this external field with the electron density $\rho({\bf r})$ of the system can be expressed by the Hamiltonian:
\begin{align}
    \hat H_{\rm Stark}=\iiint \hat\rho({\bf r})\phi_{\rm ext}({\bf r})\,{\rm d}^3{\bf r}. \label{seq:starkdef}
\end{align}
The external field $\phi_{\rm ext}$ usually acts as a perturbation to the potential of the quantum system and thus leads to energy shifts that can be obtained using the Rayleigh-Schr\"{o}dinger perturbation theory. Using Eq.\,\eqref{seq:starkdef} we calculate the Stark shift experienced by the electronic excitation between the electronic ground state $|{\rm g}\rangle$ of energy $E_{\rm g}$ and the excited state $|{\rm e}\rangle$ of energy $E_{\rm e}$ using the first-order of perturbation theory as:
\begin{align}
    \delta E_{\rm St}=\langle {\rm e}|\hat H_{\rm Stark}|{\rm e}\rangle-\langle {\rm g}|\hat H_{\rm Stark}|{\rm g}\rangle\nonumber=
    \int \Delta\rho({\bf r}) \phi_{\rm ext}({\bf r}){\rm d}^3{\bf r},
\end{align}
with $\Delta\rho({\bf r})=\rho_{\rm e}({\bf r})-\rho_{\rm g}({\bf r})$. 

We calculate the difference charge density $\Delta \rho$ by using the definition of the electronic excited state [Eq.\,\eqref{eq:excited}] and the definition of the density operator [Eq.\,\eqref{eq:densityoperator}]. The electron density of $\vert {\rm g}\rangle$ is:
\begin{align}
    \rho_{\rm g}=\langle{\rm g}\vert\hat{\rho}\vert {\rm g}\rangle=\sum_{a,\alpha}\rho_{aa}({\bf r}),
\end{align}
The electron density of the excited state can be shown to be:
\begin{align}
    \rho_{\rm e}&=\langle{\rm e}\vert\hat{\rho}\vert {\rm e}\rangle=\sum_{a,\alpha}\rho_{aa}({\bf r})\nonumber \\
    &+\sum_{ija,\alpha}C^\ast_{ia,\alpha}C_{ja,\alpha}\rho_{ij}({\bf r})-\sum_{iab,\alpha}C^\ast_{ia,\alpha}C_{ib,\alpha}\rho_{ba}({\bf r}).
\end{align}
The difference density $\Delta \rho ({\bf r})$ therefore becomes:
\begin{align}
    \Delta \rho ({\bf r})&=\rho_{\bf e}({\bf r})-\rho_{\bf g}({\bf r})\nonumber\\
    &=\sum_{ija,\alpha}C^\ast_{ia,\alpha}C_{ja,\alpha}\rho_{ij}({\bf r})-\sum_{iab,\alpha}C^\ast_{ia,\alpha}C_{ib,\alpha}\rho_{ba}({\bf r}).\label{eq:diffden}
\end{align} 

\begin{figure}[h!]
    \centering
    \includegraphics[width=0.99\linewidth]{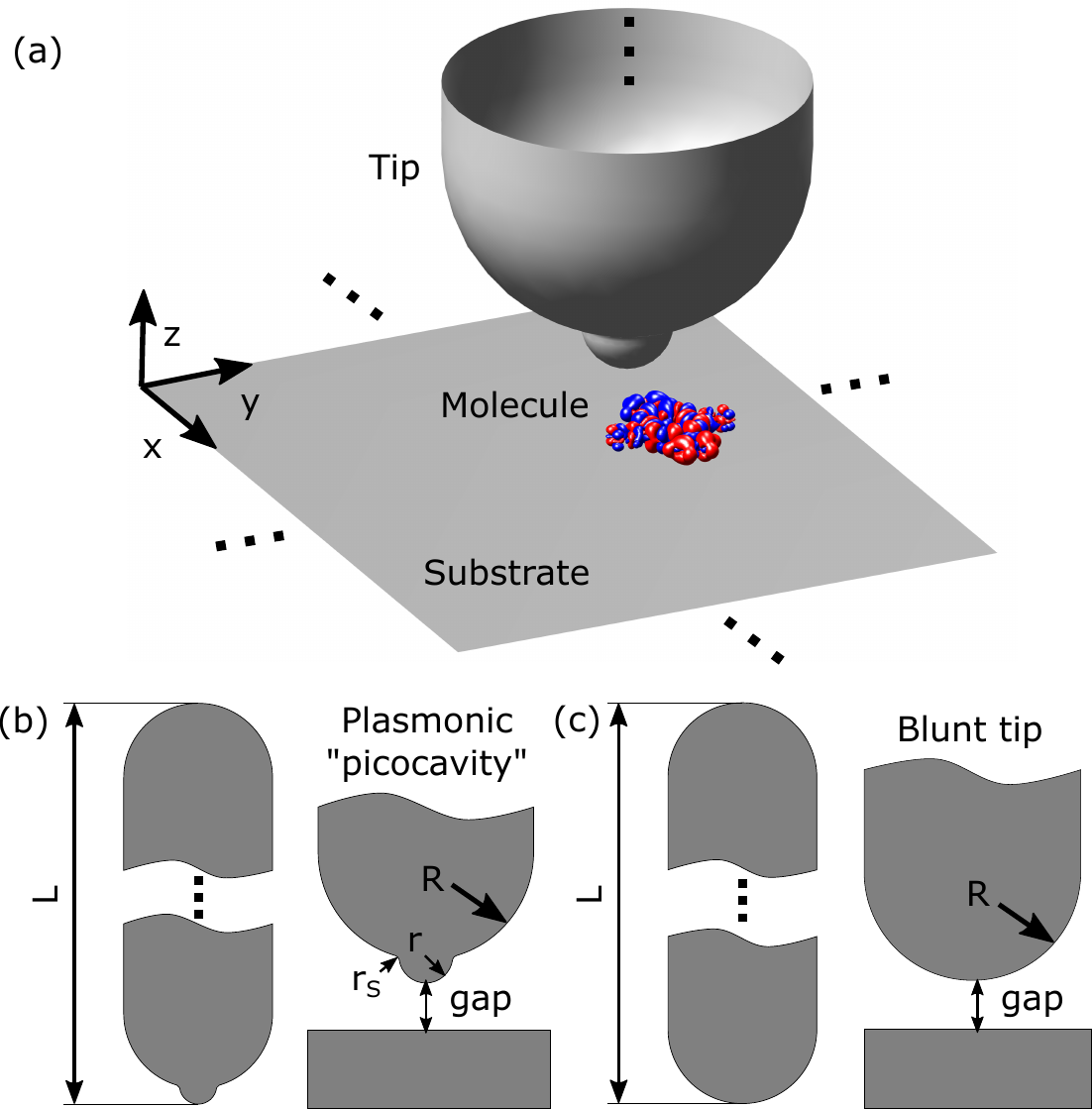}
    \caption{Geometry of the plasmonic cavities used to theoretically model the plasmonic response in STML. (a) A schematic representation of the setup containing a tip, a substrate, and a molecule (represented by the transition charge density of Q$_x$) placed into the gap between the substrate and the tip. (b) Geometry of the picocavity tip of length $L=30.5$ nm. The radius of the cylindrical capsule with hemispherical caps is $R=3$ nm. The atomistic protrusion defining the picocavity is modelled using $r=0.5$ nm, $r_{\rm s}=0.1$ nm. (c) The blunt tip is modelled as a cylindrical capsule of radius $R=3$ nm, and length $L=29.25$ nm.}
    \label{sfig:geometry}
\end{figure}

We calculate the external potential $\phi_{\rm ext}$ due to the bias voltage applied between the tip and the substrate using the finite-element method as implemented in the AC/DC module of COMSOL Multiphysics. The geometry used for the calculation is schematically shown in \Figref{sfig:geometry}. The tip is modelled as an elongated cylinder with hemispherical caps whose bottom surface additionally supports a sharp protrusion of radius $R=0.5$\,nm conservatively accounting for the atomic sharpness of the tip. 
In the practical implementation we extract the molecular orbitals $\psi_i({\bf r})$ in the form of Gaussian cube files and use them as inputs for further post processing. In the sum given by Eq.\,\eqref{eq:diffden} we consider the dominant coefficients $\sqrt{2}|C_{ia,\alpha}|>0.01$. 

\subsection{The plasmon-induced Lamb shift and Purcell effect}
We describe the plasmon-induced Lamb shift and broadening of the excitonic emission spectral lines of single molecules in STML in hybrid framework which combines the quantum properties of the molecular emitter with the classical linear-response description of the surface plasmons. The molecular emitter is represented by the position-dependent excitonic transition charge density $\rho_{\rm eg}({\bf r})$ which acts as a source of quasi-static electric field which triggers the plasmonic response in the surrounding material. The induced plasmonic potential $\phi_{\rm ind}({\bf r})$ can then be calculated using the linear-response theory, which can be expressed by defining a general response function $\chi$ of the dielectric environment as follows:
\begin{align}
\phi_{\rm ind}({\bf r}, t)=\int_{-\infty}^{t} {\rm d}\,t' \iiint {\rm d}^3 {\bf r}' \,\chi({\bf r, r}', t-t')    \tilde\rho_{\rm eg}({\bf r}', t').
\end{align}
The response function $\chi$ relates $\tilde\rho_{\rm eg}({\bf r}', t')$ at position ${\bf r}'$ and time $t'$ with the time-dependent induced plasmonic potential $\phi_{\rm ind}({\bf r}, t)$ at position ${\bf r}$ and time $t$. In the semi-classical approximation we assume that this potential is related to the mean value of the transition-charge density operator:
\begin{align}
    \tilde\rho_{\rm eg}({\bf r},t)=\rho_{\rm eg}({\bf r})\langle\sigma (t)  \rangle,
\end{align}
where $\langle \sigma(t) \rangle$ is a mean value of the oscillating lowering operator of the molecular exciton $\sigma=\lvert {\rm g}\rangle\langle{\rm e}\rvert$, with $\lvert {\rm g}\rangle$ the ground and $\lvert {\rm e}\rangle$ the excited electronic state of the molecule, respectively, and ${\rho}_{\rm eg}({\bf r})$ is the time-independent distribution of the transition charge density:
\begin{align}
    {\rho}_{\rm eg}({\bf r})=\langle{\rm g}|\hat\rho({\bf r})|{\rm e}\rangle=\sum_{ia,\alpha} C_{ia,\alpha}\rho_{ia}({\bf r}).
\end{align}
In our calculation $\rho_{\rm eg}({\bf r})$ is extracted from NWChem in the form of a Gaussian cube file. We note that the transition dipole moment of the molecule is linked to the transition-charge density via $\boldsymbol{\mu}=\iiint {\bf r}\rho_{\rm eg}({\bf r})\,{\rm d}^3{\bf r}$. Using the rotating-wave approximation we can write an effective Hamiltonian $H_{\rm ind}$ of the molecule under the influence of the induced field as:
\begin{align}
H_{\rm ind}=\iiint {\rm d}^3{\bf r}\,\phi_{\rm ind}({\bf r}, t)\rho^\ast_{\rm eg}\sigma^\dagger+{\rm H.c.}, 
\end{align}
where H.c. stands for the Hermitian conjugate. We further assume that the electronic transition in the unperturbed molecule is described by the Hamiltonian $H_{\rm mol}$:
\begin{align}
    H_{\rm mol}=\hbar\omega_{\rm eg} \sigma^\dagger\sigma, 
\end{align}
with $\omega_{\rm eg}$ being the transition frequency of the unperturbed exciton. Under the influence of $H_{\rm mol}$ the exciton of the molecule experiences time-harmonic oscillation so that $\langle\sigma(t)\rangle\propto e^{-i\omega_{\rm eg}t}$.
The mean value of the operator $\sigma$, $\langle\sigma\rangle$, evolving according to $H_{\rm tot}=H_{\rm mol}+H_{\rm ind}$ is thus governed by the linearized equation:
\begin{widetext}
\begin{align}
   \dot{ \langle\sigma(t)\rangle}=&-{\rm i}\omega_{\rm eg}\langle\sigma(t)\rangle -\frac{{\rm i}}{\hbar}\iiint {\rm d}^3{\bf r}\,\phi_{\rm ind}({\bf r},t)\rho^\ast_{\rm eg}\nonumber\\
    =&-{\rm i}\omega_{\rm eg}\langle\sigma(t)\rangle-\frac{{\rm i}}{\hbar}\int_{-\infty}^{t} {\rm d}\,t' \iiint {\rm d}^3{\bf r}\iiint {\rm d}^3 {\bf r}' \,\rho^\ast_{\rm eg}({\bf r})\chi({\bf r, r}', t-t') \rho_{\rm eg}({\bf r}')\langle\sigma(t')\rangle.
\end{align}
Next we assume that the induced potential acts in a perturbative way and induces only a minor change to the time evolution of $\langle\sigma (t)\rangle$. We therefore evaluate the time integral in the adiabatic approximation:
\begin{align}
    &\int_{-\infty}^{t} {\rm d}\,t' \iiint {\rm d}^3{\bf r}\iiint {\rm d}^3 {\bf r}' \,\rho^\ast_{\rm eg}({\bf r})\chi({\bf r, r}', t-t') \rho_{\rm eg}({\bf r}')\langle\sigma(t')\rangle\nonumber\\
    &\approx \langle\sigma(t)\rangle \int_{-\infty}^{t} {\rm d}\,t' \iiint {\rm d}^3{\bf r}\iiint {\rm d}^3 {\bf r}' \,\rho^\ast_{\rm eg}({\bf r})\chi({\bf r, r}', t-t') \rho_{\rm eg}({\bf r}')e^{{\rm i}\omega_{\rm eg}(t-t')}\nonumber\\
    &\approx \langle\sigma(t)\rangle  \iiint {\rm d}^3{\bf r}\iiint {\rm d}^3 {\bf r}' \,\rho^\ast_{\rm eg}({\bf r})\tilde\chi({\bf r, r}', \omega_{\rm eg}) \rho_{\rm eg}({\bf r}')\nonumber\\
    &=\langle\sigma(t)\rangle\iiint{\rm d}^3{\bf r}\,\rho^\ast_{\rm eg}({\bf r})\phi_{\rm ind}({\bf r},\omega_{\rm eg}).
\end{align}
Where we have defined the Fourier transform $\tilde\chi$ of the response function $\chi$, and the Fourier transform of the induced potential $\phi_{\rm ind}(\omega_{\rm eg})$ (i.e. the induced potential calculated in the frequency domain).

The time dependence of the mean value $\langle\sigma\rangle$ thus becomes:
\begin{align}
    \langle \sigma(t)\rangle \approx \langle \sigma (0)\rangle e^{-{\rm i}[\omega_{\rm eg}+f({\bf R},\omega_{\rm eg})]t},
\end{align}
with 
\begin{align}
    f({\bf R},\omega_{\rm eg})=\frac{1}{\hbar}\iiint{\rm d}^3{\bf r}\,\rho^\ast_{\rm eg}({\bf r-R})\phi_{\rm ind}({\bf r}, {\bf R},\omega_{\rm eg}).
\end{align}
The spectral response $S_{\rm e}(\omega)$ of the exciton can be obtained from the quantum regression theorem relating the dynamics of $\langle\sigma(t)\rangle$ to the dynamics of the two-time correlation function $\langle \sigma^\dagger(0)\sigma(t)\rangle\propto \langle\sigma(t)\rangle$ \cite{Breuer2003}, yielding
\begin{align}
    S_{\rm e}(\omega)\propto {\rm Re}\left\lbrace \int_0^\infty \langle\sigma^\dagger(0)\sigma(t)\rangle e^{{\rm i}\omega t}\,{\rm d}t \right\rbrace
    \propto \frac{\frac{\gamma_{\rm eg}}{{2}}}{(\omega_{\rm eg}+\delta\omega_{\rm eg}-\omega)^2+\left(\frac{\gamma_{\rm eg}}{2}\right)^2}.
\end{align}
\end{widetext}
Here the frequency shift (Lamb shift), $\delta\omega_{\rm eg}$, and line broadening (Purcell effect), $\gamma_{\rm eg}$, induced by the plasmons is related to $f({\bf R}, \omega_{\rm eg})$ as
\begin{align}
    \delta\omega_{\rm eg}={\rm Re}\{ f({\bf R},\omega_{\rm eg}) \},\label{eq:lamb}\\
    \gamma_{\rm eg}=-2{\rm Im}\{ f({\bf R},\omega_{\rm eg}) \}.\label{eq:width}
\end{align}
We use the expressions in Eq.\,\eqref{eq:lamb} and Eq.\,\eqref{eq:width} to calculate the Lamb shift and Purcell effect.

\subsection{Quasi-static calculation of the plasmonic induced electric potential}

We calculate the induced potential in the quasi-static approximation using the finite-element method implemented in the AC/DC module of COMSOL Multiphysics and describe the dielectric response of the silver tip and the substrate using the silver dielectric function obtained from \cite{JohnsonChristy1972}. In the calculations, we simplify the geometry of the tip as shown in Fig.\,\ref{sfig:geometry}. The geometry of the tip is tailored to provide a dipole plasmonic resonance along the long axis of the tip resonant with the molecular exciton. In this way, we account for the resonant plasmonic effects associated with the bright plasmonic mode of the tip that efficiently transmit the excitonic emission into the far field.
We import the excitonic transition charge density from the TD-DFT calculation and use it as a source of the classical potential generated in the geometry defined by the tip and the substrate. We impose zero-potential on the boundary of the simulation domain ensuring convergence of the results.

\subsection{Spectral response of the plasmonic cavity}\label{app:Spectral}

\begin{figure}
    \centering
    \includegraphics[width=\linewidth]{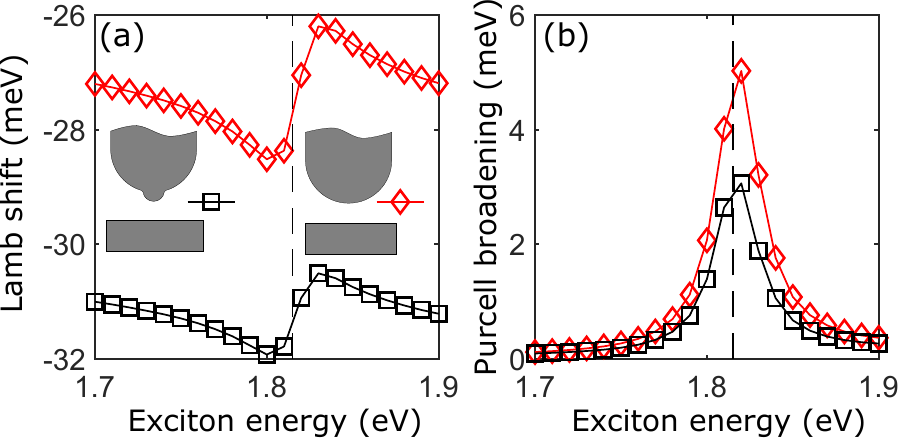}
    \caption{Spectral dependence of the Lamb shift and the Purcell effect. (a) Lamb shift calculated as a function of the exciton frequency for the picocavity (black, squares), and the blunt tip (red, diamonds). (b) Broadening of the excitonic peak due to the Purcell effect for the picocavity (black, squares), and the blunt tip (red, diamonds). The dashed black line marks the exciton energy of 1.815\,eV considered for the calculation of the spatial dependence of the Lamb shift and the Purcell effect in this work.}
    \label{sfig:spectral}
\end{figure}

We show the spectral response of the plasmonic picocavity and the cavity formed by the blunt tip and the substrate shown in \Figref{sfig:geometry}. To that end we place the molecule at the position $[x, y]=[0.69, 0.19]$\,nm and 0.5\,nm above the substrate and fix the gap size to 1\,nm. We consider that the plasmonic response is stimulated by the Q$_{x}$ exciton and treat the frequency of the exciton as a free parameter. For each frequency of the exciton we calculate the plasmonic Lamb shift and Purcell effect (broadening) and plot the results in \Figref{sfig:spectral}. Figure\,\ref{sfig:spectral}(a) shows the spectral dependence of the Lamb shift, and \Figref{sfig:spectral}(b) depicts it for the broadening of the exciton peak due to the plasmonic Purcell effect. The results for the picocavity are shown as red diamonds and for the blunt tip as black squares. A resonance feature is observed for the exciton energy around 1.815\,eV which appears as a dip followed by a peak in the Lamb-shift spectra, and has the form of a peak in the Purcell-broadening spectra. This spectral feature can be attributed to the longitudinal dipole plasmon resonance of the tip described in the main text. We further notice that the resonance feature in the Lamb-shift spectra is strongly offset by a background shift induced by higher-order plasmon modes which dominate the overall shift. Furthermore, for the exciton energy corresponding to 1.815\,eV, the shift caused resonantly by the dipole mode completely vanishes. On the other hand, the spectral response of the Purcell broadening is exclusively formed by the resonance feature caused by the interaction of the exciton with the resonant dipole plasmon mode. This supports the analysis introduced in Sec.\ref{sec:map}.  

\noindent
\section{Lorentzian fits of the Q$_{x1}$ and Q$_{x2}$ fluorescence lines.}\label{app:fits}

\begin{figure}
    \centering
    \includegraphics[width=0.8\linewidth]{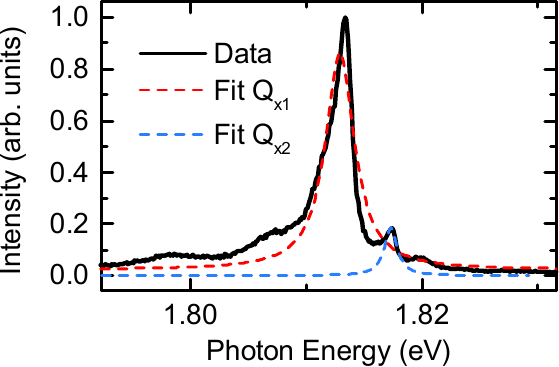}
    \caption{Typical STML spectrum acquired for a H$_{2}$Pc molecule of type 2, with Lorentzian fits adjusted to the Q$_{x1}$ and Q$_{x2}$ lines.}
    \label{sfig:lorentz}
\end{figure}

Figure \ref{sfig:lorentz} displays Lorentzian fits to the Q$_{x1}$ and Q$_{x2}$ lines. While these fits provide values for the light intensities, the peak widths and the peak positions of the two contributions that are easy to compare from one spectrum to the next, they fail to capture the asymmetric nature of the lines. As stated in the main text, this asymmetry may find its origin in different interactions (exciton-phonon, exciton-vibration, exciton-plasmon) whose spectral contributions to the STML spectra should be considered each time in a different manner. In the absence of a definitive statement regarding the peak asymmetry, we favor adjusting our spectra with simple Lorentzians that do not presume the origin of the phenomena.     

\section{Stark and Lamb Shift as a function of gap size}\label{app:StarkLamb}
\begin{figure}
    \centering
    \includegraphics[width=0.99\linewidth]{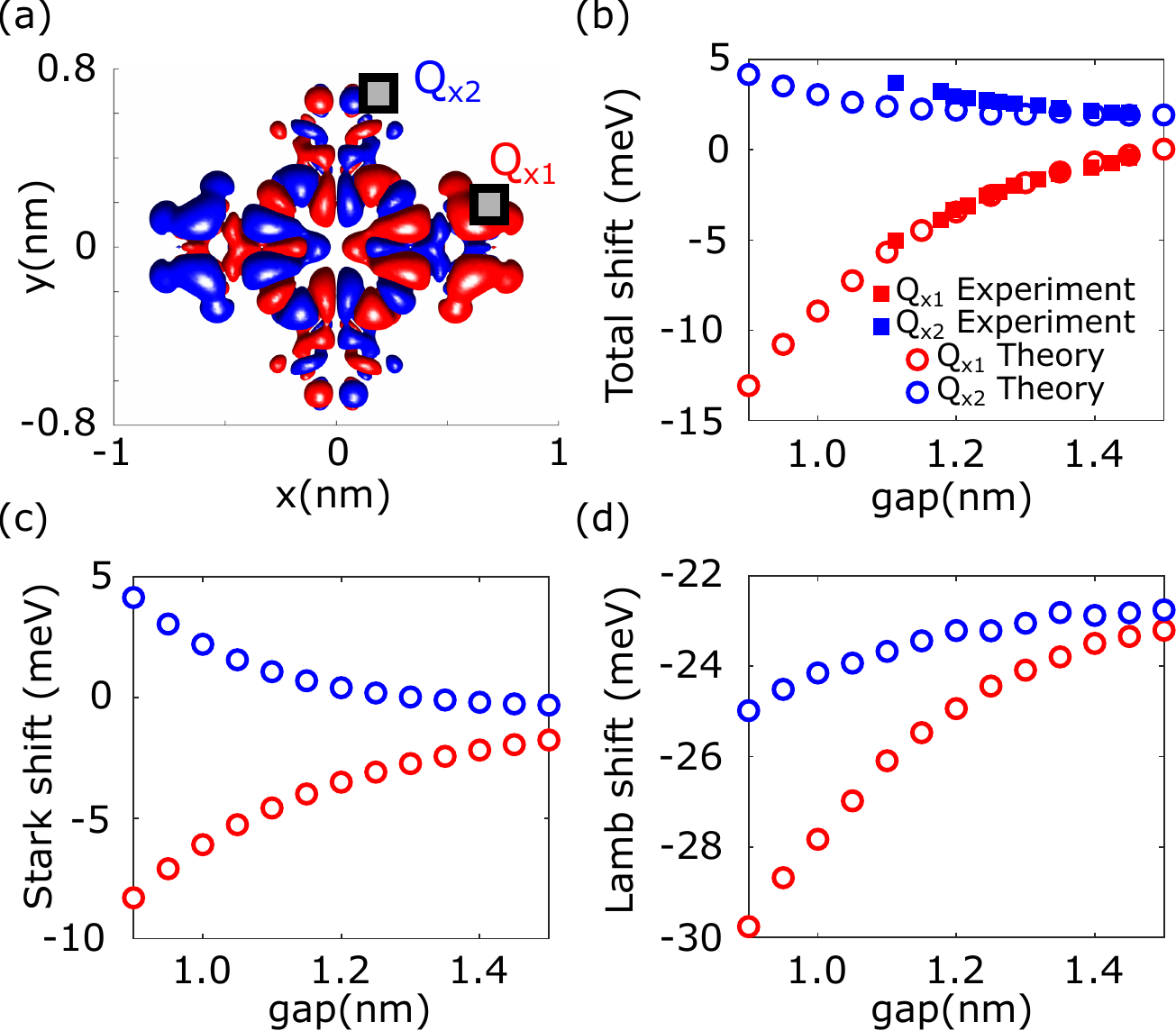}
    \caption{Excitonic peak shift as a function of tip-substrate gap. (a) Transition charge density of Q$_{x}$ with two squares marking the positions of the tip used to calculate the Stark and the Lamb shift of excitons Q$_{x1}$ and Q$_{x2}$. (b) Total peak shift calculated theoretically (circles) is displayed together with the experimental data (squares) for Q$_{x1}$ (red) and Q$_{x2}$ (blue). The origin of the energy scale is offset and corresponds to $-25$\,meV on the absolute scale. An energy of $1.815$\,eV corresponding to the energy of the unperturbed exciton on the substrate has been subtracted from the experimental data. The numerically calculated contributions to the total shift emerging from the Stark shift and from the Lamb shift are shown in (c) and (d), respectively.}
    \label{fig:approachapp}
\end{figure}

We calculate the Stark shift and the Lamb shift as a function of the gap between the tip and the substrate and show the result in Fig.\,\ref{fig:approachapp}. In the numerical calculation we account for the experimentally nonequivalent orientation of Q$_{x1}$ and Q$_{x2}$ with respect to the tip by considering two positions of the tip $[x, y]=[0.69, 0.19]$ nm (for Q$_{x1}$), and $[x,y]=[0.19, 0.69]$ nm (for Q$_{x2}$) with respect to the transition charge density of Q$_{x}$ as shown in \Figref{fig:approachapp}(a). In \Figref{fig:approachapp}(b) we show the numerically calculated total shift of the exciton as a function of the gap between the tip and the substrate for Q$_{x1}$ (red circles) and Q$_{x2}$ (blue circles).
Notice that in the theory we only determine the energy shift, but not the absolute value of the exciton energy. That is why in \Figref{fig:approachapp}(b) we offset the theoretical and experimental data. We add a constant value of $25$\,meV to the numerically calculated shift (the sum of Stark and Lamb shift). For comparison, we also display the experimentally recorded data shown in \Figref{fig2a}(d) from which we subtract the value of $1.815$\,eV, which we interpret as the energy of the exciton on the substrate, unperturbed by the tip. Finally, we measure the relative distance shown in \Figref{fig2a}(d) with respect to gap$=1.45$\,nm. The contributions to the total shift originating from the Stark effect and the Lamb shift are shown in \Figref{fig:approachapp}(c) and \Figref{fig:approachapp}(d), respectively. This decomposition shows that the Stark shift gives rise to a decrease of the energy corresponding to Q$_{x1}$ and an increase of the energy of Q$_{x2}$ as the gap size is decreased. On the other hand, the Lamb shift leads to lowering of the excitonic energy for both Q$_{x1}$ and Q$_{x2}$ with decreasing gap size, as discussed in the main text.  

\section{Probing Stark effect at a single-molecule level}\label{app:starkexperiment}
\begin{figure}
    \centering
    \includegraphics[width=1\linewidth]{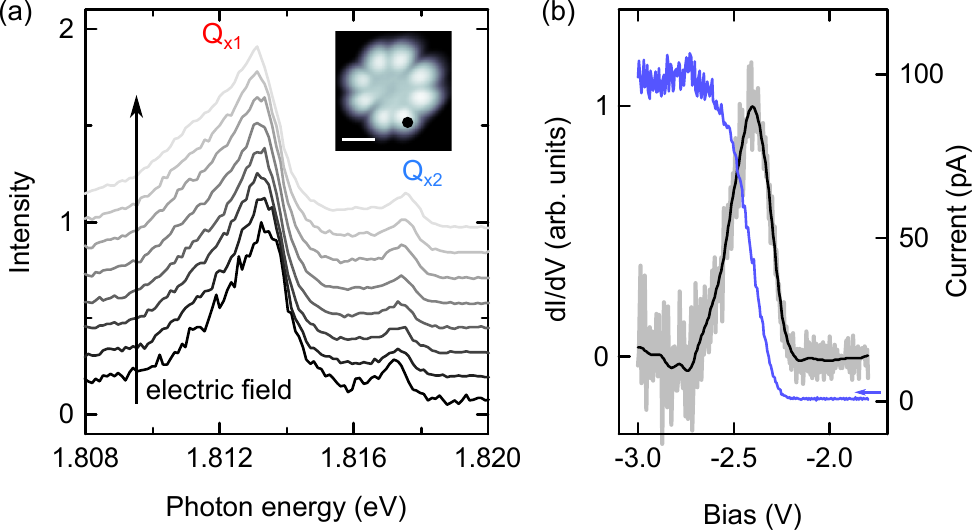}
    \caption{(a) STML spectra recorded at a fixed tip-sample distance for bias voltages varying from -2.3 V (bottom curve) to -3 V (top curve) in 0.1 V increments. Inset: STM image of the H$_{2}$Pc molecule, the measurement position is marked by a dot, $U$ = -2.5 V, $I$ = 5 pA, scale-bar: 1 nm. (b) Constant height scanning tunneling spectroscopy of the H$_{2}$Pc molecule recorded at a position marked in (a). }
    \label{sfig:Stark}
\end{figure}

In \Figref{sfig:Stark} we display normalized STML spectra acquired on the type 2 molecule discussed in \Figref{fig1}(e) for the position marked by a black dot in the STM image in the inset, at a fixed tip-molecule distance (\textit{i.e., } open feed-back loop) and for an increasing bias applied between the STM electrodes. As such, the plasmon-exciton interactions are kept constant, and the spectral shift of the Q$_{x1}$ and Q$_{x2}$ STML spectra can be readily associated to changes in the electrical field, \textit{i.e.,} the Stark effect. Here, one observes a subtle red-shift (blue-shift) of the Q$_{x1}$ (Q$_{x2}$) line with increasing electric field, an effect that is discussed in \Figref{fig2a}(e). \Figref{sfig:Stark}(b) shows a constant height scanning tunneling spectroscopy measurement and the corresponding tunneling current recorded with the same tip position as the data in (a), showing how the density of states of the molecule changes over the considered voltage range.      

\section{HRFM mapping of the Q$_{x2}$ mode of a type 1 H$_{2}$Pc molecule}\label{app:qx2}
\begin{figure}
    \centering
    \includegraphics[width=\linewidth]{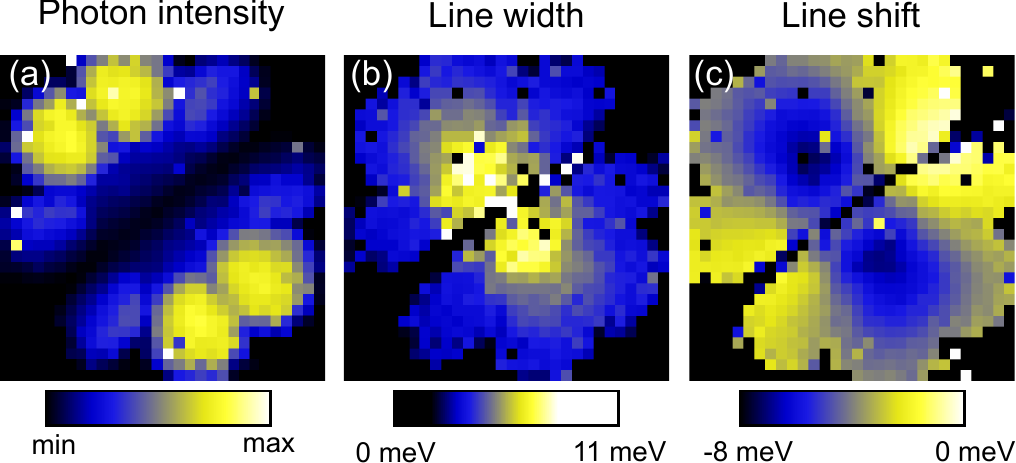}
    \caption{HRFM mapping of the Q$_{x2}$ mode of a type 1 H$_{2}$Pc molecule. (a) Photon intensity, (b) line-width and (c) line-shift maps of the Q$_{x2}$ line reconstructed from STML spectra acquired for each pixel of the map in Fig. 3a of the main text. The integration range was 1.797 - 1.837 eV. Note that, as expected, the observed spatial features are similar to the ones observed for Q$_{x1}$ in Fig. 3 of the main text, however, rotated by 90$^{\circ}$.}
    \label{sfig:Qx2}
\end{figure}

In \Figref{sfig:Qx2} we show the photon intensity, line-width and line-shift maps of the Q$_{x2}$ line (1.797 eV $\leq$ h$\nu$ $\leq$ 1.837 eV) associated to the type 1 molecule of \Figref{fig2}. As expected, these maps reveal similar patterns to the one of the Q$_{x1}$ spectral line but tilted by 90$^\circ$. One also notices that the line-width maxima and line-shift minima are located slightly closer to the center of the molecule compared to the Q$_{x1}$ maps, a behaviour that we tentatively associate to an environmental effect linked to different molecular dipole-substrate configurations.    

\newpage

\bibliography{HRFM}

\begin{thebibliography}{58}%
\makeatletter
\providecommand \@ifxundefined [1]{%
 \@ifx{#1\undefined}
}%
\providecommand \@ifnum [1]{%
 \ifnum #1\expandafter \@firstoftwo
 \else \expandafter \@secondoftwo
 \fi
}%
\providecommand \@ifx [1]{%
 \ifx #1\expandafter \@firstoftwo
 \else \expandafter \@secondoftwo
 \fi
}%
\providecommand \natexlab [1]{#1}%
\providecommand \enquote  [1]{``#1''}%
\providecommand \bibnamefont  [1]{#1}%
\providecommand \bibfnamefont [1]{#1}%
\providecommand \citenamefont [1]{#1}%
\providecommand \href@noop [0]{\@secondoftwo}%
\providecommand \href [0]{\begingroup \@sanitize@url \@href}%
\providecommand \@href[1]{\@@startlink{#1}\@@href}%
\providecommand \@@href[1]{\endgroup#1\@@endlink}%
\providecommand \@sanitize@url [0]{\catcode `\\12\catcode `\$12\catcode
  `\&12\catcode `\#12\catcode `\^12\catcode `\_12\catcode `\%12\relax}%
\providecommand \@@startlink[1]{}%
\providecommand \@@endlink[0]{}%
\providecommand \url  [0]{\begingroup\@sanitize@url \@url }%
\providecommand \@url [1]{\endgroup\@href {#1}{\urlprefix }}%
\providecommand \urlprefix  [0]{URL }%
\providecommand \Eprint [0]{\href }%
\providecommand \doibase [0]{https://doi.org/}%
\providecommand \selectlanguage [0]{\@gobble}%
\providecommand \bibinfo  [0]{\@secondoftwo}%
\providecommand \bibfield  [0]{\@secondoftwo}%
\providecommand \translation [1]{[#1]}%
\providecommand \BibitemOpen [0]{}%
\providecommand \bibitemStop [0]{}%
\providecommand \bibitemNoStop [0]{.\EOS\space}%
\providecommand \EOS [0]{\spacefactor3000\relax}%
\providecommand \BibitemShut  [1]{\csname bibitem#1\endcsname}%
\let\auto@bib@innerbib\@empty
%</preamble>
\bibitem [{\citenamefont {Moerner}\ and\ \citenamefont
  {Kador}(1989)}]{Moerner1989}%
  \BibitemOpen
  \bibfield  {author} {\bibinfo {author} {\bibfnamefont {W.~E.}\ \bibnamefont
  {Moerner}}\ and\ \bibinfo {author} {\bibfnamefont {L.}~\bibnamefont
  {Kador}},\ }\bibfield  {title} {\bibinfo {title} {Optical detection and
  spectroscopy of single molecules in a solid},\ }\href
  {https://doi.org/10.1103/PhysRevLett.62.2535} {\bibfield  {journal} {\bibinfo
   {journal} {Physical Review Letters}\ }\textbf {\bibinfo {volume} {62}},\
  \bibinfo {pages} {2535} (\bibinfo {year} {1989})}\BibitemShut {NoStop}%
\bibitem [{\citenamefont {Orrit}\ and\ \citenamefont
  {Bernard}(1990)}]{Orrit1990}%
  \BibitemOpen
  \bibfield  {author} {\bibinfo {author} {\bibfnamefont {M.}~\bibnamefont
  {Orrit}}\ and\ \bibinfo {author} {\bibfnamefont {J.}~\bibnamefont
  {Bernard}},\ }\bibfield  {title} {\bibinfo {title} {Single pentacene
  molecules detected by fluorescence excitation in a p-terphenyl crystal},\
  }\href {https://doi.org/10.1103/PhysRevLett.65.2716} {\bibfield  {journal}
  {\bibinfo  {journal} {Physical Review Letters}\ }\textbf {\bibinfo {volume}
  {65}},\ \bibinfo {pages} {2716} (\bibinfo {year} {1990})}\BibitemShut
  {NoStop}%
\bibitem [{\citenamefont {Huang}\ \emph {et~al.}(2009)\citenamefont {Huang},
  \citenamefont {Bates},\ and\ \citenamefont {Zhuang}}]{Huang2009}%
  \BibitemOpen
  \bibfield  {author} {\bibinfo {author} {\bibfnamefont {B.}~\bibnamefont
  {Huang}}, \bibinfo {author} {\bibfnamefont {M.}~\bibnamefont {Bates}},\ and\
  \bibinfo {author} {\bibfnamefont {X.}~\bibnamefont {Zhuang}},\ }\bibfield
  {title} {\bibinfo {title} {Super-resolution fluorescence microscopy},\ }\href
  {https://doi.org/10.1146/annurev.biochem.77.061906.092014} {\bibfield
  {journal} {\bibinfo  {journal} {Annu. Rev. Biochem.}\ }\textbf {\bibinfo
  {volume} {78}},\ \bibinfo {pages} {993} (\bibinfo {year} {2009})}\BibitemShut
  {NoStop}%
\bibitem [{\citenamefont {Hell}\ \emph {et~al.}(2015)\citenamefont {Hell},
  \citenamefont {Sahl}, \citenamefont {Bates}, \citenamefont {Zhuang},
  \citenamefont {Heintzmann}, \citenamefont {Booth}, \citenamefont
  {Bewersdorf}, \citenamefont {Shtengel}, \citenamefont {Hess}, \citenamefont
  {Tinnefeld}, \citenamefont {Honigmann}, \citenamefont {Jakobs}, \citenamefont
  {Testa}, \citenamefont {Cognet}, \citenamefont {Lounis}, \citenamefont
  {Ewers}, \citenamefont {Davis}, \citenamefont {Eggeling}, \citenamefont
  {Klenerman}, \citenamefont {Willig}, \citenamefont {Vicidomini},
  \citenamefont {Castello}, \citenamefont {Diaspro},\ and\ \citenamefont
  {Cordes}}]{Hell2015}%
  \BibitemOpen
  \bibfield  {author} {\bibinfo {author} {\bibfnamefont {S.~W.}\ \bibnamefont
  {Hell}}, \bibinfo {author} {\bibfnamefont {S.~J.}\ \bibnamefont {Sahl}},
  \bibinfo {author} {\bibfnamefont {M.}~\bibnamefont {Bates}}, \bibinfo
  {author} {\bibfnamefont {X.}~\bibnamefont {Zhuang}}, \bibinfo {author}
  {\bibfnamefont {R.}~\bibnamefont {Heintzmann}}, \bibinfo {author}
  {\bibfnamefont {M.~J.}\ \bibnamefont {Booth}}, \bibinfo {author}
  {\bibfnamefont {J.}~\bibnamefont {Bewersdorf}}, \bibinfo {author}
  {\bibfnamefont {G.}~\bibnamefont {Shtengel}}, \bibinfo {author}
  {\bibfnamefont {H.}~\bibnamefont {Hess}}, \bibinfo {author} {\bibfnamefont
  {P.}~\bibnamefont {Tinnefeld}}, \bibinfo {author} {\bibfnamefont
  {A.}~\bibnamefont {Honigmann}}, \bibinfo {author} {\bibfnamefont
  {S.}~\bibnamefont {Jakobs}}, \bibinfo {author} {\bibfnamefont
  {I.}~\bibnamefont {Testa}}, \bibinfo {author} {\bibfnamefont
  {L.}~\bibnamefont {Cognet}}, \bibinfo {author} {\bibfnamefont
  {B.}~\bibnamefont {Lounis}}, \bibinfo {author} {\bibfnamefont
  {H.}~\bibnamefont {Ewers}}, \bibinfo {author} {\bibfnamefont {S.~J.}\
  \bibnamefont {Davis}}, \bibinfo {author} {\bibfnamefont {C.}~\bibnamefont
  {Eggeling}}, \bibinfo {author} {\bibfnamefont {D.}~\bibnamefont {Klenerman}},
  \bibinfo {author} {\bibfnamefont {K.~I.}\ \bibnamefont {Willig}}, \bibinfo
  {author} {\bibfnamefont {G.}~\bibnamefont {Vicidomini}}, \bibinfo {author}
  {\bibfnamefont {M.}~\bibnamefont {Castello}}, \bibinfo {author}
  {\bibfnamefont {A.}~\bibnamefont {Diaspro}},\ and\ \bibinfo {author}
  {\bibfnamefont {T.}~\bibnamefont {Cordes}},\ }\bibfield  {title} {\bibinfo
  {title} {The 2015 super-resolution microscopy roadmap},\ }\href
  {https://doi.org/10.1088/0022-3727/48/44/443001} {\bibfield  {journal}
  {\bibinfo  {journal} {Journal of Physics D: Applied Physics}\ }\textbf
  {\bibinfo {volume} {48}},\ \bibinfo {pages} {443001} (\bibinfo {year}
  {2015})}\BibitemShut {NoStop}%
\bibitem [{\citenamefont {Lounis}\ and\ \citenamefont
  {Moerner}(2000)}]{Lounis2000}%
  \BibitemOpen
  \bibfield  {author} {\bibinfo {author} {\bibfnamefont {B.}~\bibnamefont
  {Lounis}}\ and\ \bibinfo {author} {\bibfnamefont {W.~E.}\ \bibnamefont
  {Moerner}},\ }\bibfield  {title} {\bibinfo {title} {Single photons on demand
  from a single molecule at room temperature},\ }\href
  {https://doi.org/10.1038/35035032} {\bibfield  {journal} {\bibinfo  {journal}
  {Nature}\ }\textbf {\bibinfo {volume} {407}},\ \bibinfo {pages} {491}
  (\bibinfo {year} {2000})}\BibitemShut {NoStop}%
\bibitem [{\citenamefont {Toninelli}\ \emph {et~al.}(2020)\citenamefont
  {Toninelli} \emph {et~al.}}]{toninelli2020single}%
  \BibitemOpen
  \bibfield  {author} {\bibinfo {author} {\bibfnamefont {C.}~\bibnamefont
  {Toninelli}} \emph {et~al.},\ }\href@noop {} {} (\bibinfo {year} {2020}),\
  \Eprint {https://arxiv.org/abs/2011.05059} {arXiv:2011.05059 [quant-ph]}
  \BibitemShut {NoStop}%
\bibitem [{\citenamefont {Purcell}\ \emph {et~al.}(1946)\citenamefont
  {Purcell}, \citenamefont {Torrey},\ and\ \citenamefont
  {Pound}}]{purcell1946spontaneous}%
  \BibitemOpen
  \bibfield  {author} {\bibinfo {author} {\bibfnamefont {E.~M.}\ \bibnamefont
  {Purcell}}, \bibinfo {author} {\bibfnamefont {H.~C.}\ \bibnamefont
  {Torrey}},\ and\ \bibinfo {author} {\bibfnamefont {R.~V.}\ \bibnamefont
  {Pound}},\ }\bibfield  {title} {\bibinfo {title} {Resonance absorption by
  nuclear magnetic moments in a solid},\ }\href
  {https://doi.org/10.1103/PhysRev.69.37} {\bibfield  {journal} {\bibinfo
  {journal} {Phys. Rev.}\ }\textbf {\bibinfo {volume} {69}},\ \bibinfo {pages}
  {37} (\bibinfo {year} {1946})}\BibitemShut {NoStop}%
\bibitem [{\citenamefont {Novotny}\ and\ \citenamefont
  {Hecht}(2006)}]{novotny_hecht_2006}%
  \BibitemOpen
  \bibfield  {author} {\bibinfo {author} {\bibfnamefont {L.}~\bibnamefont
  {Novotny}}\ and\ \bibinfo {author} {\bibfnamefont {B.}~\bibnamefont
  {Hecht}},\ }\href {https://doi.org/10.1017/CBO9780511813535} {\emph {\bibinfo
  {title} {Principles of Nano-Optics}}}\ (\bibinfo  {publisher} {Cambridge
  University Press},\ \bibinfo {year} {2006})\BibitemShut {NoStop}%
\bibitem [{\citenamefont {Lamb}\ and\ \citenamefont
  {Retherford}(1947)}]{Lamb1947}%
  \BibitemOpen
  \bibfield  {author} {\bibinfo {author} {\bibfnamefont {W.~E.}\ \bibnamefont
  {Lamb}}\ and\ \bibinfo {author} {\bibfnamefont {R.~C.}\ \bibnamefont
  {Retherford}},\ }\bibfield  {title} {\bibinfo {title} {Fine structure of the
  hydrogen atom by a microwave method},\ }\href
  {https://doi.org/10.1103/PhysRev.72.241} {\bibfield  {journal} {\bibinfo
  {journal} {Phys. Rev.}\ }\textbf {\bibinfo {volume} {72}},\ \bibinfo {pages}
  {241} (\bibinfo {year} {1947})}\BibitemShut {NoStop}%
\bibitem [{\citenamefont {Bethe}(1947)}]{bethe1947}%
  \BibitemOpen
  \bibfield  {author} {\bibinfo {author} {\bibfnamefont {H.~A.}\ \bibnamefont
  {Bethe}},\ }\bibfield  {title} {\bibinfo {title} {The electromagnetic shift
  of energy levels},\ }\href {https://doi.org/10.1103/PhysRev.72.339}
  {\bibfield  {journal} {\bibinfo  {journal} {Phys. Rev.}\ }\textbf {\bibinfo
  {volume} {72}},\ \bibinfo {pages} {339} (\bibinfo {year} {1947})}\BibitemShut
  {NoStop}%
\bibitem [{\citenamefont {Bethe}\ \emph {et~al.}(1950)\citenamefont {Bethe},
  \citenamefont {Brown},\ and\ \citenamefont {Stehn}}]{bethe1950}%
  \BibitemOpen
  \bibfield  {author} {\bibinfo {author} {\bibfnamefont {H.~A.}\ \bibnamefont
  {Bethe}}, \bibinfo {author} {\bibfnamefont {L.~M.}\ \bibnamefont {Brown}},\
  and\ \bibinfo {author} {\bibfnamefont {J.~R.}\ \bibnamefont {Stehn}},\
  }\bibfield  {title} {\bibinfo {title} {Numerical value of the lamb shift},\
  }\href {https://doi.org/10.1103/PhysRev.77.370} {\bibfield  {journal}
  {\bibinfo  {journal} {Phys. Rev.}\ }\textbf {\bibinfo {volume} {77}},\
  \bibinfo {pages} {370} (\bibinfo {year} {1950})}\BibitemShut {NoStop}%
\bibitem [{\citenamefont {Power}(1966)}]{Power1966}%
  \BibitemOpen
  \bibfield  {author} {\bibinfo {author} {\bibfnamefont {E.~A.}\ \bibnamefont
  {Power}},\ }\bibfield  {title} {\bibinfo {title} {Zero-point energy and the
  lamb shift},\ }\href {https://doi.org/10.1119/1.1973082} {\bibfield
  {journal} {\bibinfo  {journal} {American Journal of Physics}\ }\textbf
  {\bibinfo {volume} {34}},\ \bibinfo {pages} {516} (\bibinfo {year}
  {1966})}\BibitemShut {NoStop}%
\bibitem [{\citenamefont {Stark}(1914)}]{stark1914}%
  \BibitemOpen
  \bibfield  {author} {\bibinfo {author} {\bibfnamefont {J.}~\bibnamefont
  {Stark}},\ }\bibfield  {title} {\bibinfo {title} {Beobachtungen über den
  effekt des elektrischen feldes auf spektrallinien. i. quereffekt},\ }\href
  {https://doi.org/https://doi.org/10.1002/andp.19143480702} {\bibfield
  {journal} {\bibinfo  {journal} {Annalen der Physik}\ }\textbf {\bibinfo
  {volume} {348}},\ \bibinfo {pages} {965} (\bibinfo {year}
  {1914})}\BibitemShut {NoStop}%
\bibitem [{\citenamefont {Krems}(2018)}]{starkeffect2018}%
  \BibitemOpen
  \bibfield  {author} {\bibinfo {author} {\bibfnamefont {R.~V.}\ \bibnamefont
  {Krems}},\ }\bibinfo {title} {{DC} stark effect},\ in\ \href
  {https://doi.org/https://doi.org/10.1002/9781119382638.ch2} {\emph {\bibinfo
  {booktitle} {Molecules in Electromagnetic Fields}}}\ (\bibinfo  {publisher}
  {John Wiley \& Sons, Ltd},\ \bibinfo {year} {2018})\ Chap.~\bibinfo {chapter}
  {2}, pp.\ \bibinfo {pages} {35--58}\BibitemShut {NoStop}%
\bibitem [{\citenamefont {Betzig}\ and\ \citenamefont
  {Chichester}(1993)}]{Betzig1993}%
  \BibitemOpen
  \bibfield  {author} {\bibinfo {author} {\bibfnamefont {E.}~\bibnamefont
  {Betzig}}\ and\ \bibinfo {author} {\bibfnamefont {R.~J.}\ \bibnamefont
  {Chichester}},\ }\bibfield  {title} {\bibinfo {title} {Single molecules
  observed by near-field scanning optical microscopy},\ }\href
  {https://doi.org/10.1126/science.262.5138.1422} {\bibfield  {journal}
  {\bibinfo  {journal} {Science}\ }\textbf {\bibinfo {volume} {262}},\ \bibinfo
  {pages} {1422} (\bibinfo {year} {1993})}\BibitemShut {NoStop}%
\bibitem [{\citenamefont {Keilmann}\ and\ \citenamefont
  {Hillenbrand}(2004)}]{Richards2004ssnom}%
  \BibitemOpen
  \bibfield  {author} {\bibinfo {author} {\bibfnamefont {F.}~\bibnamefont
  {Keilmann}}\ and\ \bibinfo {author} {\bibfnamefont {R.}~\bibnamefont
  {Hillenbrand}},\ }\bibfield  {title} {\bibinfo {title} {Near-field microscopy
  by elastic light scattering from a tip},\ }\href
  {https://doi.org/10.1098/rsta.2003.1347} {\bibfield  {journal} {\bibinfo
  {journal} {Phil. Trans. R. Soc. A}\ }\textbf {\bibinfo {volume} {362}},\
  \bibinfo {pages} {787} (\bibinfo {year} {2004})}\BibitemShut {NoStop}%
\bibitem [{\citenamefont {Frey}\ \emph {et~al.}(2004)\citenamefont {Frey},
  \citenamefont {Witt}, \citenamefont {Felderer},\ and\ \citenamefont
  {Guckenberger}}]{frey2004highresolution}%
  \BibitemOpen
  \bibfield  {author} {\bibinfo {author} {\bibfnamefont {H.~G.}\ \bibnamefont
  {Frey}}, \bibinfo {author} {\bibfnamefont {S.}~\bibnamefont {Witt}}, \bibinfo
  {author} {\bibfnamefont {K.}~\bibnamefont {Felderer}},\ and\ \bibinfo
  {author} {\bibfnamefont {R.}~\bibnamefont {Guckenberger}},\ }\bibfield
  {title} {\bibinfo {title} {High-resolution imaging of single fluorescent
  molecules with the optical near-field of a metal tip},\ }\href
  {https://doi.org/10.1103/PhysRevLett.93.200801} {\bibfield  {journal}
  {\bibinfo  {journal} {Phys. Rev. Lett.}\ }\textbf {\bibinfo {volume} {93}},\
  \bibinfo {pages} {200801} (\bibinfo {year} {2004})}\BibitemShut {NoStop}%
\bibitem [{\citenamefont {Novotny}\ and\ \citenamefont
  {Stranick}(2006)}]{Novotny2006}%
  \BibitemOpen
  \bibfield  {author} {\bibinfo {author} {\bibfnamefont {L.}~\bibnamefont
  {Novotny}}\ and\ \bibinfo {author} {\bibfnamefont {S.~J.}\ \bibnamefont
  {Stranick}},\ }\bibfield  {title} {\bibinfo {title} {Near-field optical
  microscopy and spectroscopy with pointed probes},\ }\href
  {https://doi.org/10.1146/annurev.physchem.56.092503.141236} {\bibfield
  {journal} {\bibinfo  {journal} {Annu. Rev. Phys. Chem.}\ }\textbf {\bibinfo
  {volume} {57}},\ \bibinfo {pages} {303} (\bibinfo {year} {2006})}\BibitemShut
  {NoStop}%
\bibitem [{\citenamefont {K\"uhn}\ \emph {et~al.}(2006)\citenamefont {K\"uhn},
  \citenamefont {H\aa{}kanson}, \citenamefont {Rogobete},\ and\ \citenamefont
  {Sandoghdar}}]{Kuhn2006enhancement}%
  \BibitemOpen
  \bibfield  {author} {\bibinfo {author} {\bibfnamefont {S.}~\bibnamefont
  {K\"uhn}}, \bibinfo {author} {\bibfnamefont {U.}~\bibnamefont
  {H\aa{}kanson}}, \bibinfo {author} {\bibfnamefont {L.}~\bibnamefont
  {Rogobete}},\ and\ \bibinfo {author} {\bibfnamefont {V.}~\bibnamefont
  {Sandoghdar}},\ }\bibfield  {title} {\bibinfo {title} {Enhancement of
  single-molecule fluorescence using a gold nanoparticle as an optical
  nanoantenna},\ }\href {https://doi.org/10.1103/PhysRevLett.97.017402}
  {\bibfield  {journal} {\bibinfo  {journal} {Phys. Rev. Lett.}\ }\textbf
  {\bibinfo {volume} {97}},\ \bibinfo {pages} {017402} (\bibinfo {year}
  {2006})}\BibitemShut {NoStop}%
\bibitem [{\citenamefont {Hartschuh}(2008)}]{hartschuh2008tipenhanced}%
  \BibitemOpen
  \bibfield  {author} {\bibinfo {author} {\bibfnamefont {A.}~\bibnamefont
  {Hartschuh}},\ }\bibfield  {title} {\bibinfo {title} {Tip-enhanced near-field
  optical microscopy},\ }\href {https://doi.org/10.1002/anie.200801605}
  {\bibfield  {journal} {\bibinfo  {journal} {Angew. Chem. Int. Ed. Engl.}\
  }\textbf {\bibinfo {volume} {47}},\ \bibinfo {pages} {8178} (\bibinfo {year}
  {2008})}\BibitemShut {NoStop}%
\bibitem [{\citenamefont {Mauser}\ and\ \citenamefont
  {Hartschuh}(2014)}]{mauser2014tipenhanced}%
  \BibitemOpen
  \bibfield  {author} {\bibinfo {author} {\bibfnamefont {N.}~\bibnamefont
  {Mauser}}\ and\ \bibinfo {author} {\bibfnamefont {A.}~\bibnamefont
  {Hartschuh}},\ }\bibfield  {title} {\bibinfo {title} {Tip-enhanced near-field
  optical microscopy},\ }\href {https://doi.org/10.1039/C3CS60258C} {\bibfield
  {journal} {\bibinfo  {journal} {Chem. Soc. Rev.}\ }\textbf {\bibinfo {volume}
  {43}},\ \bibinfo {pages} {1248} (\bibinfo {year} {2014})}\BibitemShut
  {NoStop}%
\bibitem [{\citenamefont {Park}\ \emph {et~al.}(2018)\citenamefont {Park},
  \citenamefont {Jiang}, \citenamefont {Clark}, \citenamefont {Xu},\ and\
  \citenamefont {Raschke}}]{Park2018radiative}%
  \BibitemOpen
  \bibfield  {author} {\bibinfo {author} {\bibfnamefont {K.-D.}\ \bibnamefont
  {Park}}, \bibinfo {author} {\bibfnamefont {T.}~\bibnamefont {Jiang}},
  \bibinfo {author} {\bibfnamefont {G.}~\bibnamefont {Clark}}, \bibinfo
  {author} {\bibfnamefont {X.}~\bibnamefont {Xu}},\ and\ \bibinfo {author}
  {\bibfnamefont {M.~B.}\ \bibnamefont {Raschke}},\ }\bibfield  {title}
  {\bibinfo {title} {Radiative control of dark excitons at room temperature by
  nano-optical antenna-tip purcell effect},\ }\href
  {https://doi.org/10.1038/s41565-017-0003-0} {\bibfield  {journal} {\bibinfo
  {journal} {Nature Nanotech.}\ }\textbf {\bibinfo {volume} {13}},\ \bibinfo
  {pages} {59} (\bibinfo {year} {2018})}\BibitemShut {NoStop}%
\bibitem [{\citenamefont {Zhang}\ \emph {et~al.}(2013)\citenamefont {Zhang},
  \citenamefont {Zhang}, \citenamefont {Dong}, \citenamefont {Jiang},
  \citenamefont {Zhang}, \citenamefont {Chen}, \citenamefont {Zhang},
  \citenamefont {Liao}, \citenamefont {Aizpurua}, \citenamefont {Luo},
  \citenamefont {Yang},\ and\ \citenamefont {Hou}}]{Zhang2013}%
  \BibitemOpen
  \bibfield  {author} {\bibinfo {author} {\bibfnamefont {R.}~\bibnamefont
  {Zhang}}, \bibinfo {author} {\bibfnamefont {Y.}~\bibnamefont {Zhang}},
  \bibinfo {author} {\bibfnamefont {Z.~C.}\ \bibnamefont {Dong}}, \bibinfo
  {author} {\bibfnamefont {S.}~\bibnamefont {Jiang}}, \bibinfo {author}
  {\bibfnamefont {C.}~\bibnamefont {Zhang}}, \bibinfo {author} {\bibfnamefont
  {L.~G.}\ \bibnamefont {Chen}}, \bibinfo {author} {\bibfnamefont
  {L.}~\bibnamefont {Zhang}}, \bibinfo {author} {\bibfnamefont
  {Y.}~\bibnamefont {Liao}}, \bibinfo {author} {\bibfnamefont {J.}~\bibnamefont
  {Aizpurua}}, \bibinfo {author} {\bibfnamefont {Y.}~\bibnamefont {Luo}},
  \bibinfo {author} {\bibfnamefont {J.~L.}\ \bibnamefont {Yang}},\ and\
  \bibinfo {author} {\bibfnamefont {J.~G.}\ \bibnamefont {Hou}},\ }\bibfield
  {title} {\bibinfo {title} {Chemical mapping of a single molecule by
  plasmon-enhanced raman scattering},\ }\href
  {https://doi.org/10.1038/nature12151} {\bibfield  {journal} {\bibinfo
  {journal} {Nature}\ }\textbf {\bibinfo {volume} {498}},\ \bibinfo {pages}
  {82} (\bibinfo {year} {2013})}\BibitemShut {NoStop}%
\bibitem [{\citenamefont {Lee}\ \emph {et~al.}(2019)\citenamefont {Lee},
  \citenamefont {Crampton}, \citenamefont {Tallarida},\ and\ \citenamefont
  {Apkarian}}]{Lee2019}%
  \BibitemOpen
  \bibfield  {author} {\bibinfo {author} {\bibfnamefont {J.}~\bibnamefont
  {Lee}}, \bibinfo {author} {\bibfnamefont {K.~T.}\ \bibnamefont {Crampton}},
  \bibinfo {author} {\bibfnamefont {N.}~\bibnamefont {Tallarida}},\ and\
  \bibinfo {author} {\bibfnamefont {V.~A.}\ \bibnamefont {Apkarian}},\
  }\bibfield  {title} {\bibinfo {title} {Visualizing vibrational normal modes
  of a single molecule with atomically confined light},\ }\href
  {https://doi.org/10.1038/s41586-019-1059-9} {\bibfield  {journal} {\bibinfo
  {journal} {Nature}\ }\textbf {\bibinfo {volume} {568}},\ \bibinfo {pages}
  {78} (\bibinfo {year} {2019})}\BibitemShut {NoStop}%
\bibitem [{\citenamefont {Yang}\ \emph {et~al.}(2020)\citenamefont {Yang},
  \citenamefont {Chen}, \citenamefont {Ghafoor}, \citenamefont {Zhang},
  \citenamefont {Zhang}, \citenamefont {Zhang}, \citenamefont {Luo},
  \citenamefont {Yang}, \citenamefont {Sandoghdar}, \citenamefont {Aizpurua},
  \citenamefont {Dong},\ and\ \citenamefont {Hou}}]{Yang2020subnanometer}%
  \BibitemOpen
  \bibfield  {author} {\bibinfo {author} {\bibfnamefont {B.}~\bibnamefont
  {Yang}}, \bibinfo {author} {\bibfnamefont {G.}~\bibnamefont {Chen}}, \bibinfo
  {author} {\bibfnamefont {A.}~\bibnamefont {Ghafoor}}, \bibinfo {author}
  {\bibfnamefont {Y.}~\bibnamefont {Zhang}}, \bibinfo {author} {\bibfnamefont
  {Y.}~\bibnamefont {Zhang}}, \bibinfo {author} {\bibfnamefont
  {Y.}~\bibnamefont {Zhang}}, \bibinfo {author} {\bibfnamefont
  {Y.}~\bibnamefont {Luo}}, \bibinfo {author} {\bibfnamefont {J.}~\bibnamefont
  {Yang}}, \bibinfo {author} {\bibfnamefont {V.}~\bibnamefont {Sandoghdar}},
  \bibinfo {author} {\bibfnamefont {J.}~\bibnamefont {Aizpurua}}, \bibinfo
  {author} {\bibfnamefont {Z.}~\bibnamefont {Dong}},\ and\ \bibinfo {author}
  {\bibfnamefont {J.~G.}\ \bibnamefont {Hou}},\ }\bibfield  {title} {\bibinfo
  {title} {Sub-nanometre resolution in single-molecule photoluminescence
  imaging},\ }\href {https://doi.org/10.1038/s41566-020-0677-y} {\bibfield
  {journal} {\bibinfo  {journal} {Nat. Photonics}\ }\textbf {\bibinfo {volume}
  {14}},\ \bibinfo {pages} {693} (\bibinfo {year} {2020})}\BibitemShut
  {NoStop}%
\bibitem [{\citenamefont {Qiu}\ \emph {et~al.}(2003)\citenamefont {Qiu},
  \citenamefont {Nazin},\ and\ \citenamefont {Ho}}]{Qiu2003}%
  \BibitemOpen
  \bibfield  {author} {\bibinfo {author} {\bibfnamefont {X.~H.}\ \bibnamefont
  {Qiu}}, \bibinfo {author} {\bibfnamefont {G.~V.}\ \bibnamefont {Nazin}},\
  and\ \bibinfo {author} {\bibfnamefont {W.}~\bibnamefont {Ho}},\ }\bibfield
  {title} {\bibinfo {title} {Vibrationally resolved fluorescence excited with
  submolecular precision},\ }\href {https://doi.org/10.1126/science.1078675}
  {\bibfield  {journal} {\bibinfo  {journal} {Science}\ }\textbf {\bibinfo
  {volume} {299}},\ \bibinfo {pages} {542} (\bibinfo {year}
  {2003})}\BibitemShut {NoStop}%
\bibitem [{\citenamefont {Chen}\ \emph {et~al.}(2010)\citenamefont {Chen},
  \citenamefont {Chu}, \citenamefont {Bobisch}, \citenamefont {Mills},\ and\
  \citenamefont {Ho}}]{Chen2010}%
  \BibitemOpen
  \bibfield  {author} {\bibinfo {author} {\bibfnamefont {C.}~\bibnamefont
  {Chen}}, \bibinfo {author} {\bibfnamefont {P.}~\bibnamefont {Chu}}, \bibinfo
  {author} {\bibfnamefont {C.~A.}\ \bibnamefont {Bobisch}}, \bibinfo {author}
  {\bibfnamefont {D.~L.}\ \bibnamefont {Mills}},\ and\ \bibinfo {author}
  {\bibfnamefont {W.}~\bibnamefont {Ho}},\ }\bibfield  {title} {\bibinfo
  {title} {Viewing the interior of a single molecule: Vibronically resolved
  photon imaging at submolecular resolution},\ }\href
  {https://doi.org/10.1103/PhysRevLett.105.217402} {\bibfield  {journal}
  {\bibinfo  {journal} {Phys. Rev. Lett.}\ }\textbf {\bibinfo {volume} {105}},\
  \bibinfo {pages} {217402} (\bibinfo {year} {2010})}\BibitemShut {NoStop}%
\bibitem [{\citenamefont {Chong}\ \emph
  {et~al.}(2016{\natexlab{a}})\citenamefont {Chong}, \citenamefont {Reecht},
  \citenamefont {Bulou}, \citenamefont {Boeglin}, \citenamefont {Scheurer},
  \citenamefont {Mathevet},\ and\ \citenamefont {Schull}}]{Chong2016}%
  \BibitemOpen
  \bibfield  {author} {\bibinfo {author} {\bibfnamefont {M.~C.}\ \bibnamefont
  {Chong}}, \bibinfo {author} {\bibfnamefont {G.}~\bibnamefont {Reecht}},
  \bibinfo {author} {\bibfnamefont {H.}~\bibnamefont {Bulou}}, \bibinfo
  {author} {\bibfnamefont {A.}~\bibnamefont {Boeglin}}, \bibinfo {author}
  {\bibfnamefont {F.}~\bibnamefont {Scheurer}}, \bibinfo {author}
  {\bibfnamefont {F.}~\bibnamefont {Mathevet}},\ and\ \bibinfo {author}
  {\bibfnamefont {G.}~\bibnamefont {Schull}},\ }\bibfield  {title} {\bibinfo
  {title} {Narrow-line single-molecule transducer between electronic circuits
  and surface plasmons},\ }\href
  {https://doi.org/10.1103/PhysRevLett.116.036802} {\bibfield  {journal}
  {\bibinfo  {journal} {Phys. Rev. Lett.}\ }\textbf {\bibinfo {volume} {116}},\
  \bibinfo {pages} {036802} (\bibinfo {year} {2016}{\natexlab{a}})}\BibitemShut
  {NoStop}%
\bibitem [{\citenamefont {Chong}\ \emph
  {et~al.}(2016{\natexlab{b}})\citenamefont {Chong}, \citenamefont
  {{Sosa-Vargas}}, \citenamefont {Bulou}, \citenamefont {Boeglin},
  \citenamefont {Scheurer}, \citenamefont {Mathevet},\ and\ \citenamefont
  {Schull}}]{Chong2016ordinary}%
  \BibitemOpen
  \bibfield  {author} {\bibinfo {author} {\bibfnamefont {M.~C.}\ \bibnamefont
  {Chong}}, \bibinfo {author} {\bibfnamefont {L.}~\bibnamefont
  {{Sosa-Vargas}}}, \bibinfo {author} {\bibfnamefont {H.}~\bibnamefont
  {Bulou}}, \bibinfo {author} {\bibfnamefont {A.}~\bibnamefont {Boeglin}},
  \bibinfo {author} {\bibfnamefont {F.}~\bibnamefont {Scheurer}}, \bibinfo
  {author} {\bibfnamefont {F.}~\bibnamefont {Mathevet}},\ and\ \bibinfo
  {author} {\bibfnamefont {G.}~\bibnamefont {Schull}},\ }\bibfield  {title}
  {\bibinfo {title} {Ordinary and {{Hot Electroluminescence}} from
  {{Single}}-{{Molecule Devices}}: {{Controlling}} the {{Emission Color}} by
  {{Chemical Engineering}}},\ }\href
  {https://doi.org/10.1021/acs.nanolett.6b02997} {\bibfield  {journal}
  {\bibinfo  {journal} {Nano Letters}\ }\textbf {\bibinfo {volume} {16}},\
  \bibinfo {pages} {6480} (\bibinfo {year} {2016}{\natexlab{b}})}\BibitemShut
  {NoStop}%
\bibitem [{\citenamefont {Zhang}\ \emph {et~al.}(2016)\citenamefont {Zhang},
  \citenamefont {Luo}, \citenamefont {Zhang}, \citenamefont {Yu}, \citenamefont
  {Kuang}, \citenamefont {Zhang}, \citenamefont {Meng}, \citenamefont {Luo},
  \citenamefont {Yang}, \citenamefont {Dong},\ and\ \citenamefont
  {Hou}}]{Zhang2016}%
  \BibitemOpen
  \bibfield  {author} {\bibinfo {author} {\bibfnamefont {Y.}~\bibnamefont
  {Zhang}}, \bibinfo {author} {\bibfnamefont {Y.}~\bibnamefont {Luo}}, \bibinfo
  {author} {\bibfnamefont {Y.}~\bibnamefont {Zhang}}, \bibinfo {author}
  {\bibfnamefont {Y.-J.}\ \bibnamefont {Yu}}, \bibinfo {author} {\bibfnamefont
  {Y.-M.}\ \bibnamefont {Kuang}}, \bibinfo {author} {\bibfnamefont
  {L.}~\bibnamefont {Zhang}}, \bibinfo {author} {\bibfnamefont {Q.-S.}\
  \bibnamefont {Meng}}, \bibinfo {author} {\bibfnamefont {Y.}~\bibnamefont
  {Luo}}, \bibinfo {author} {\bibfnamefont {J.-L.}\ \bibnamefont {Yang}},
  \bibinfo {author} {\bibfnamefont {Z.-C.}\ \bibnamefont {Dong}},\ and\
  \bibinfo {author} {\bibfnamefont {J.~G.}\ \bibnamefont {Hou}},\ }\bibfield
  {title} {\bibinfo {title} {Visualizing coherent intermolecular dipole-dipole
  coupling in real space},\ }\href {https://doi.org/10.1038/nature17428}
  {\bibfield  {journal} {\bibinfo  {journal} {Nature}\ }\textbf {\bibinfo
  {volume} {531}},\ \bibinfo {pages} {623} (\bibinfo {year}
  {2016})}\BibitemShut {NoStop}%
\bibitem [{\citenamefont {Imada}\ \emph {et~al.}(2016)\citenamefont {Imada},
  \citenamefont {Miwa}, \citenamefont {{Imai-Imada}}, \citenamefont {Kawahara},
  \citenamefont {Kimura},\ and\ \citenamefont {Kim}}]{Imada2016}%
  \BibitemOpen
  \bibfield  {author} {\bibinfo {author} {\bibfnamefont {H.}~\bibnamefont
  {Imada}}, \bibinfo {author} {\bibfnamefont {K.}~\bibnamefont {Miwa}},
  \bibinfo {author} {\bibfnamefont {M.}~\bibnamefont {{Imai-Imada}}}, \bibinfo
  {author} {\bibfnamefont {S.}~\bibnamefont {Kawahara}}, \bibinfo {author}
  {\bibfnamefont {K.}~\bibnamefont {Kimura}},\ and\ \bibinfo {author}
  {\bibfnamefont {Y.}~\bibnamefont {Kim}},\ }\bibfield  {title} {\bibinfo
  {title} {Real-space investigation of energy transfer in heterogeneous
  molecular dimers},\ }\href {https://doi.org/10.1038/nature19765} {\bibfield
  {journal} {\bibinfo  {journal} {Nature}\ }\textbf {\bibinfo {volume} {538}},\
  \bibinfo {pages} {364} (\bibinfo {year} {2016})}\BibitemShut {NoStop}%
\bibitem [{\citenamefont {Gro{\ss}e}\ \emph {et~al.}(2017)\citenamefont
  {Gro{\ss}e}, \citenamefont {Merino}, \citenamefont {Ros{\l}awska},
  \citenamefont {Gunnarsson}, \citenamefont {Kuhnke},\ and\ \citenamefont
  {Kern}}]{Grosse2017}%
  \BibitemOpen
  \bibfield  {author} {\bibinfo {author} {\bibfnamefont {C.}~\bibnamefont
  {Gro{\ss}e}}, \bibinfo {author} {\bibfnamefont {P.}~\bibnamefont {Merino}},
  \bibinfo {author} {\bibfnamefont {A.}~\bibnamefont {Ros{\l}awska}}, \bibinfo
  {author} {\bibfnamefont {O.}~\bibnamefont {Gunnarsson}}, \bibinfo {author}
  {\bibfnamefont {K.}~\bibnamefont {Kuhnke}},\ and\ \bibinfo {author}
  {\bibfnamefont {K.}~\bibnamefont {Kern}},\ }\bibfield  {title} {\bibinfo
  {title} {Submolecular {{Electroluminescence Mapping}} of {{Organic
  Semiconductors}}},\ }\href {https://doi.org/10.1021/acsnano.6b08471}
  {\bibfield  {journal} {\bibinfo  {journal} {ACS Nano}\ }\textbf {\bibinfo
  {volume} {11}},\ \bibinfo {pages} {1230} (\bibinfo {year}
  {2017})}\BibitemShut {NoStop}%
\bibitem [{\citenamefont {Doppagne}\ \emph {et~al.}(2017)\citenamefont
  {Doppagne}, \citenamefont {Chong}, \citenamefont {Lorchat}, \citenamefont
  {Berciaud}, \citenamefont {Romeo}, \citenamefont {Bulou}, \citenamefont
  {Boeglin}, \citenamefont {Scheurer},\ and\ \citenamefont
  {Schull}}]{Doppagne2017}%
  \BibitemOpen
  \bibfield  {author} {\bibinfo {author} {\bibfnamefont {B.}~\bibnamefont
  {Doppagne}}, \bibinfo {author} {\bibfnamefont {M.~C.}\ \bibnamefont {Chong}},
  \bibinfo {author} {\bibfnamefont {E.}~\bibnamefont {Lorchat}}, \bibinfo
  {author} {\bibfnamefont {S.}~\bibnamefont {Berciaud}}, \bibinfo {author}
  {\bibfnamefont {M.}~\bibnamefont {Romeo}}, \bibinfo {author} {\bibfnamefont
  {H.}~\bibnamefont {Bulou}}, \bibinfo {author} {\bibfnamefont
  {A.}~\bibnamefont {Boeglin}}, \bibinfo {author} {\bibfnamefont
  {F.}~\bibnamefont {Scheurer}},\ and\ \bibinfo {author} {\bibfnamefont
  {G.}~\bibnamefont {Schull}},\ }\bibfield  {title} {\bibinfo {title} {Vibronic
  {{Spectroscopy}} with {{Submolecular Resolution}} from {{STM}}-{{Induced
  Electroluminescence}}},\ }\href
  {https://doi.org/10.1103/PhysRevLett.118.127401} {\bibfield  {journal}
  {\bibinfo  {journal} {Phys. Rev. Lett.}\ }\textbf {\bibinfo {volume} {118}},\
  \bibinfo {pages} {127401} (\bibinfo {year} {2017})}\BibitemShut {NoStop}%
\bibitem [{\citenamefont {Imada}\ \emph {et~al.}(2017)\citenamefont {Imada},
  \citenamefont {Miwa}, \citenamefont {{Imai-Imada}}, \citenamefont {Kawahara},
  \citenamefont {Kimura},\ and\ \citenamefont {Kim}}]{Imada2017}%
  \BibitemOpen
  \bibfield  {author} {\bibinfo {author} {\bibfnamefont {H.}~\bibnamefont
  {Imada}}, \bibinfo {author} {\bibfnamefont {K.}~\bibnamefont {Miwa}},
  \bibinfo {author} {\bibfnamefont {M.}~\bibnamefont {{Imai-Imada}}}, \bibinfo
  {author} {\bibfnamefont {S.}~\bibnamefont {Kawahara}}, \bibinfo {author}
  {\bibfnamefont {K.}~\bibnamefont {Kimura}},\ and\ \bibinfo {author}
  {\bibfnamefont {Y.}~\bibnamefont {Kim}},\ }\bibfield  {title} {\bibinfo
  {title} {Single-{{Molecule Investigation}} of {{Energy Dynamics}} in a
  {{Coupled Plasmon}}-{{Exciton System}}},\ }\href
  {https://doi.org/10.1103/PhysRevLett.119.013901} {\bibfield  {journal}
  {\bibinfo  {journal} {Phys. Rev. Lett.}\ }\textbf {\bibinfo {volume} {119}},\
  \bibinfo {pages} {013901} (\bibinfo {year} {2017})}\BibitemShut {NoStop}%
\bibitem [{\citenamefont {Zhang}\ \emph
  {et~al.}(2017{\natexlab{a}})\citenamefont {Zhang}, \citenamefont {Meng},
  \citenamefont {Zhang}, \citenamefont {Luo}, \citenamefont {Yu}, \citenamefont
  {Yang}, \citenamefont {Zhang}, \citenamefont {Esteban}, \citenamefont
  {Aizpurua}, \citenamefont {Luo}, \citenamefont {Yang}, \citenamefont {Dong},\
  and\ \citenamefont {Hou}}]{Zhang2017_vtomas}%
  \BibitemOpen
  \bibfield  {author} {\bibinfo {author} {\bibfnamefont {Y.}~\bibnamefont
  {Zhang}}, \bibinfo {author} {\bibfnamefont {Q.-S.}\ \bibnamefont {Meng}},
  \bibinfo {author} {\bibfnamefont {L.}~\bibnamefont {Zhang}}, \bibinfo
  {author} {\bibfnamefont {Y.}~\bibnamefont {Luo}}, \bibinfo {author}
  {\bibfnamefont {Y.-J.}\ \bibnamefont {Yu}}, \bibinfo {author} {\bibfnamefont
  {B.}~\bibnamefont {Yang}}, \bibinfo {author} {\bibfnamefont {Y.}~\bibnamefont
  {Zhang}}, \bibinfo {author} {\bibfnamefont {R.}~\bibnamefont {Esteban}},
  \bibinfo {author} {\bibfnamefont {J.}~\bibnamefont {Aizpurua}}, \bibinfo
  {author} {\bibfnamefont {Y.}~\bibnamefont {Luo}}, \bibinfo {author}
  {\bibfnamefont {J.-L.}\ \bibnamefont {Yang}}, \bibinfo {author}
  {\bibfnamefont {Z.-C.}\ \bibnamefont {Dong}},\ and\ \bibinfo {author}
  {\bibfnamefont {J.~G.}\ \bibnamefont {Hou}},\ }\bibfield  {title} {\bibinfo
  {title} {Sub-nanometre control of the coherent interaction between a single
  molecule and a plasmonic nanocavity},\ }\href
  {https://doi.org/10.1038/ncomms15225} {\bibfield  {journal} {\bibinfo
  {journal} {Nat. Commun.}\ }\textbf {\bibinfo {volume} {8}},\ \bibinfo {pages}
  {15225} (\bibinfo {year} {2017}{\natexlab{a}})}\BibitemShut {NoStop}%
\bibitem [{\citenamefont {Zhang}\ \emph
  {et~al.}(2017{\natexlab{b}})\citenamefont {Zhang}, \citenamefont {Yu},
  \citenamefont {Chen}, \citenamefont {Luo}, \citenamefont {Yang},
  \citenamefont {Kong}, \citenamefont {Chen}, \citenamefont {Zhang},
  \citenamefont {Zhang}, \citenamefont {Luo}, \citenamefont {Yang},
  \citenamefont {Dong},\ and\ \citenamefont {Hou}}]{Zhang2017_a}%
  \BibitemOpen
  \bibfield  {author} {\bibinfo {author} {\bibfnamefont {L.}~\bibnamefont
  {Zhang}}, \bibinfo {author} {\bibfnamefont {Y.-J.}\ \bibnamefont {Yu}},
  \bibinfo {author} {\bibfnamefont {L.-G.}\ \bibnamefont {Chen}}, \bibinfo
  {author} {\bibfnamefont {Y.}~\bibnamefont {Luo}}, \bibinfo {author}
  {\bibfnamefont {B.}~\bibnamefont {Yang}}, \bibinfo {author} {\bibfnamefont
  {F.-F.}\ \bibnamefont {Kong}}, \bibinfo {author} {\bibfnamefont
  {G.}~\bibnamefont {Chen}}, \bibinfo {author} {\bibfnamefont {Y.}~\bibnamefont
  {Zhang}}, \bibinfo {author} {\bibfnamefont {Q.}~\bibnamefont {Zhang}},
  \bibinfo {author} {\bibfnamefont {Y.}~\bibnamefont {Luo}}, \bibinfo {author}
  {\bibfnamefont {J.-L.}\ \bibnamefont {Yang}}, \bibinfo {author}
  {\bibfnamefont {Z.-C.}\ \bibnamefont {Dong}},\ and\ \bibinfo {author}
  {\bibfnamefont {J.~G.}\ \bibnamefont {Hou}},\ }\bibfield  {title} {\bibinfo
  {title} {Electrically driven single-photon emission from an isolated single
  molecule},\ }\href {https://doi.org/10.1038/s41467-017-00681-7} {\bibfield
  {journal} {\bibinfo  {journal} {Nature Communications}\ }\textbf {\bibinfo
  {volume} {8}},\ \bibinfo {pages} {580} (\bibinfo {year}
  {2017}{\natexlab{b}})}\BibitemShut {NoStop}%
\bibitem [{\citenamefont {Doppagne}\ \emph {et~al.}(2018)\citenamefont
  {Doppagne}, \citenamefont {Chong}, \citenamefont {Bulou}, \citenamefont
  {Boeglin}, \citenamefont {Scheurer},\ and\ \citenamefont
  {Schull}}]{Doppagne2018}%
  \BibitemOpen
  \bibfield  {author} {\bibinfo {author} {\bibfnamefont {B.}~\bibnamefont
  {Doppagne}}, \bibinfo {author} {\bibfnamefont {M.~C.}\ \bibnamefont {Chong}},
  \bibinfo {author} {\bibfnamefont {H.}~\bibnamefont {Bulou}}, \bibinfo
  {author} {\bibfnamefont {A.}~\bibnamefont {Boeglin}}, \bibinfo {author}
  {\bibfnamefont {F.}~\bibnamefont {Scheurer}},\ and\ \bibinfo {author}
  {\bibfnamefont {G.}~\bibnamefont {Schull}},\ }\bibfield  {title} {\bibinfo
  {title} {Electrofluorochromism at the single-molecule level},\ }\href
  {https://doi.org/10.1126/science.aat1603} {\bibfield  {journal} {\bibinfo
  {journal} {Science}\ }\textbf {\bibinfo {volume} {361}},\ \bibinfo {pages}
  {251} (\bibinfo {year} {2018})}\BibitemShut {NoStop}%
\bibitem [{\citenamefont {Dole{\v z}al}\ \emph {et~al.}(2019)\citenamefont
  {Dole{\v z}al}, \citenamefont {Merino}, \citenamefont {Redondo},
  \citenamefont {Ondi{\v c}}, \citenamefont {Cahl{\'i}k},\ and\ \citenamefont
  {{\v S}vec}}]{Dolezal2019}%
  \BibitemOpen
  \bibfield  {author} {\bibinfo {author} {\bibfnamefont {J.}~\bibnamefont
  {Dole{\v z}al}}, \bibinfo {author} {\bibfnamefont {P.}~\bibnamefont
  {Merino}}, \bibinfo {author} {\bibfnamefont {J.}~\bibnamefont {Redondo}},
  \bibinfo {author} {\bibfnamefont {L.}~\bibnamefont {Ondi{\v c}}}, \bibinfo
  {author} {\bibfnamefont {A.}~\bibnamefont {Cahl{\'i}k}},\ and\ \bibinfo
  {author} {\bibfnamefont {M.}~\bibnamefont {{\v S}vec}},\ }\bibfield  {title}
  {\bibinfo {title} {Charge {{Carrier Injection Electroluminescence}} with
  {{CO}}-{{Functionalized Tips}} on {{Single Molecular Emitters}}},\ }\href
  {https://doi.org/10.1021/acs.nanolett.9b03180} {\bibfield  {journal}
  {\bibinfo  {journal} {Nano Lett.}\ }\textbf {\bibinfo {volume} {19}},\
  \bibinfo {pages} {8605} (\bibinfo {year} {2019})}\BibitemShut {NoStop}%
\bibitem [{\citenamefont {Doppagne}\ \emph {et~al.}(2020)\citenamefont
  {Doppagne}, \citenamefont {Neuman}, \citenamefont {{Soria-Martinez}},
  \citenamefont {L{\'o}pez}, \citenamefont {Bulou}, \citenamefont {Romeo},
  \citenamefont {Berciaud}, \citenamefont {Scheurer}, \citenamefont
  {Aizpurua},\ and\ \citenamefont {Schull}}]{Doppagne2020}%
  \BibitemOpen
  \bibfield  {author} {\bibinfo {author} {\bibfnamefont {B.}~\bibnamefont
  {Doppagne}}, \bibinfo {author} {\bibfnamefont {T.}~\bibnamefont {Neuman}},
  \bibinfo {author} {\bibfnamefont {R.}~\bibnamefont {{Soria-Martinez}}},
  \bibinfo {author} {\bibfnamefont {L.~E.~P.}\ \bibnamefont {L{\'o}pez}},
  \bibinfo {author} {\bibfnamefont {H.}~\bibnamefont {Bulou}}, \bibinfo
  {author} {\bibfnamefont {M.}~\bibnamefont {Romeo}}, \bibinfo {author}
  {\bibfnamefont {S.}~\bibnamefont {Berciaud}}, \bibinfo {author}
  {\bibfnamefont {F.}~\bibnamefont {Scheurer}}, \bibinfo {author}
  {\bibfnamefont {J.}~\bibnamefont {Aizpurua}},\ and\ \bibinfo {author}
  {\bibfnamefont {G.}~\bibnamefont {Schull}},\ }\bibfield  {title} {\bibinfo
  {title} {Single-molecule tautomerization tracking through space- and
  time-resolved fluorescence spectroscopy},\ }\href
  {https://doi.org/10.1038/s41565-019-0620-x} {\bibfield  {journal} {\bibinfo
  {journal} {Nat. Nanotechnol.}\ }\textbf {\bibinfo {volume} {15}},\ \bibinfo
  {pages} {207} (\bibinfo {year} {2020})}\BibitemShut {NoStop}%
\bibitem [{\citenamefont {Rai}\ \emph {et~al.}(2020)\citenamefont {Rai},
  \citenamefont {Gerhard}, \citenamefont {Sun}, \citenamefont {Holzer},
  \citenamefont {Rep{\"a}n}, \citenamefont {Krsti{\'c}}, \citenamefont {Yang},
  \citenamefont {Wegener}, \citenamefont {Rockstuhl},\ and\ \citenamefont
  {Wulfhekel}}]{Rai2020}%
  \BibitemOpen
  \bibfield  {author} {\bibinfo {author} {\bibfnamefont {V.}~\bibnamefont
  {Rai}}, \bibinfo {author} {\bibfnamefont {L.}~\bibnamefont {Gerhard}},
  \bibinfo {author} {\bibfnamefont {Q.}~\bibnamefont {Sun}}, \bibinfo {author}
  {\bibfnamefont {C.}~\bibnamefont {Holzer}}, \bibinfo {author} {\bibfnamefont
  {T.}~\bibnamefont {Rep{\"a}n}}, \bibinfo {author} {\bibfnamefont
  {M.}~\bibnamefont {Krsti{\'c}}}, \bibinfo {author} {\bibfnamefont
  {L.}~\bibnamefont {Yang}}, \bibinfo {author} {\bibfnamefont {M.}~\bibnamefont
  {Wegener}}, \bibinfo {author} {\bibfnamefont {C.}~\bibnamefont {Rockstuhl}},\
  and\ \bibinfo {author} {\bibfnamefont {W.}~\bibnamefont {Wulfhekel}},\
  }\bibfield  {title} {\bibinfo {title} {Boosting {{Light Emission}} from
  {{Single Hydrogen Phthalocyanine Molecules}} by {{Charging}}},\ }\href
  {https://doi.org/10.1021/acs.nanolett.0c03121} {\bibfield  {journal}
  {\bibinfo  {journal} {Nano Lett.}\ }\textbf {\bibinfo {volume} {20}},\
  \bibinfo {pages} {7600} (\bibinfo {year} {2020})}\BibitemShut {NoStop}%
\bibitem [{\citenamefont {Kuhnke}\ \emph {et~al.}(2017)\citenamefont {Kuhnke},
  \citenamefont {Turkowski}, \citenamefont {Kabakchiev}, \citenamefont {Lutz},
  \citenamefont {Rahman},\ and\ \citenamefont {Kern}}]{Kuhnke2017a}%
  \BibitemOpen
  \bibfield  {author} {\bibinfo {author} {\bibfnamefont {K.}~\bibnamefont
  {Kuhnke}}, \bibinfo {author} {\bibfnamefont {V.}~\bibnamefont {Turkowski}},
  \bibinfo {author} {\bibfnamefont {A.}~\bibnamefont {Kabakchiev}}, \bibinfo
  {author} {\bibfnamefont {T.}~\bibnamefont {Lutz}}, \bibinfo {author}
  {\bibfnamefont {T.~S.}\ \bibnamefont {Rahman}},\ and\ \bibinfo {author}
  {\bibfnamefont {K.}~\bibnamefont {Kern}},\ }\bibfield  {title} {\bibinfo
  {title} {Pentacene {{Excitons}} in {{Strong Electric Fields}}},\ }\href
  {https://doi.org/10.1002/cphc.201701174} {\bibfield  {journal} {\bibinfo
  {journal} {Chem. Phys. Chem.}\ }\textbf {\bibinfo {volume} {19}},\ \bibinfo
  {pages} {277} (\bibinfo {year} {2017})}\BibitemShut {NoStop}%
\bibitem [{\citenamefont {Benz}\ \emph {et~al.}(2016)\citenamefont {Benz},
  \citenamefont {Schmidt}, \citenamefont {Dreismann}, \citenamefont
  {Chikkaraddy}, \citenamefont {Zhang}, \citenamefont {Demetriadou},
  \citenamefont {Carnegie}, \citenamefont {Ohadi}, \citenamefont {de~Nijs},
  \citenamefont {Esteban}, \citenamefont {Aizpurua},\ and\ \citenamefont
  {Baumberg}}]{Benz2016}%
  \BibitemOpen
  \bibfield  {author} {\bibinfo {author} {\bibfnamefont {F.}~\bibnamefont
  {Benz}}, \bibinfo {author} {\bibfnamefont {M.~K.}\ \bibnamefont {Schmidt}},
  \bibinfo {author} {\bibfnamefont {A.}~\bibnamefont {Dreismann}}, \bibinfo
  {author} {\bibfnamefont {R.}~\bibnamefont {Chikkaraddy}}, \bibinfo {author}
  {\bibfnamefont {Y.}~\bibnamefont {Zhang}}, \bibinfo {author} {\bibfnamefont
  {A.}~\bibnamefont {Demetriadou}}, \bibinfo {author} {\bibfnamefont
  {C.}~\bibnamefont {Carnegie}}, \bibinfo {author} {\bibfnamefont
  {H.}~\bibnamefont {Ohadi}}, \bibinfo {author} {\bibfnamefont
  {B.}~\bibnamefont {de~Nijs}}, \bibinfo {author} {\bibfnamefont
  {R.}~\bibnamefont {Esteban}}, \bibinfo {author} {\bibfnamefont
  {J.}~\bibnamefont {Aizpurua}},\ and\ \bibinfo {author} {\bibfnamefont
  {J.~J.}\ \bibnamefont {Baumberg}},\ }\bibfield  {title} {\bibinfo {title}
  {Single-molecule optomechanics in ``picocavities''},\ }\href
  {https://doi.org/10.1126/science.aah5243} {\bibfield  {journal} {\bibinfo
  {journal} {Science}\ }\textbf {\bibinfo {volume} {354}},\ \bibinfo {pages}
  {726} (\bibinfo {year} {2016})}\BibitemShut {NoStop}%
\bibitem [{\citenamefont {Barbry}\ \emph {et~al.}(2015)\citenamefont {Barbry},
  \citenamefont {Koval}, \citenamefont {Marchesin}, \citenamefont {Esteban},
  \citenamefont {Borisov}, \citenamefont {Aizpurua},\ and\ \citenamefont
  {{S{\'a}nchez-Portal}}}]{Barbry2015}%
  \BibitemOpen
  \bibfield  {author} {\bibinfo {author} {\bibfnamefont {M.}~\bibnamefont
  {Barbry}}, \bibinfo {author} {\bibfnamefont {P.}~\bibnamefont {Koval}},
  \bibinfo {author} {\bibfnamefont {F.}~\bibnamefont {Marchesin}}, \bibinfo
  {author} {\bibfnamefont {R.}~\bibnamefont {Esteban}}, \bibinfo {author}
  {\bibfnamefont {A.~G.}\ \bibnamefont {Borisov}}, \bibinfo {author}
  {\bibfnamefont {J.}~\bibnamefont {Aizpurua}},\ and\ \bibinfo {author}
  {\bibfnamefont {D.}~\bibnamefont {{S{\'a}nchez-Portal}}},\ }\bibfield
  {title} {\bibinfo {title} {Atomistic {{Near}}-{{Field Nanoplasmonics}}:
  {{Reaching Atomic}}-{{Scale Resolution}} in {{Nanooptics}}},\ }\href
  {https://doi.org/10.1021/acs.nanolett.5b00759} {\bibfield  {journal}
  {\bibinfo  {journal} {Nano Letters}\ }\textbf {\bibinfo {volume} {15}},\
  \bibinfo {pages} {3410} (\bibinfo {year} {2015})}\BibitemShut {NoStop}%
\bibitem [{\citenamefont {Urbieta}\ \emph {et~al.}(2018)\citenamefont
  {Urbieta}, \citenamefont {Barbry}, \citenamefont {Zhang}, \citenamefont
  {Koval}, \citenamefont {S\'{a}nchez-Portal}, \citenamefont {Zabala},\ and\
  \citenamefont {Aizpurua}}]{urbieta2018}%
  \BibitemOpen
  \bibfield  {author} {\bibinfo {author} {\bibfnamefont {M.}~\bibnamefont
  {Urbieta}}, \bibinfo {author} {\bibfnamefont {M.}~\bibnamefont {Barbry}},
  \bibinfo {author} {\bibfnamefont {Y.}~\bibnamefont {Zhang}}, \bibinfo
  {author} {\bibfnamefont {P.}~\bibnamefont {Koval}}, \bibinfo {author}
  {\bibfnamefont {D.}~\bibnamefont {S\'{a}nchez-Portal}}, \bibinfo {author}
  {\bibfnamefont {N.}~\bibnamefont {Zabala}},\ and\ \bibinfo {author}
  {\bibfnamefont {J.}~\bibnamefont {Aizpurua}},\ }\bibfield  {title} {\bibinfo
  {title} {Atomic-scale lightning rod effect in plasmonic picocavities: A
  classical view to a quantum effect},\ }\href@noop {} {\bibfield  {journal}
  {\bibinfo  {journal} {ACS Nano}\ }\textbf {\bibinfo {volume} {12}},\ \bibinfo
  {pages} {585} (\bibinfo {year} {2018})}\BibitemShut {NoStop}%
\bibitem [{\citenamefont {Rossi}\ \emph {et~al.}(2019)\citenamefont {Rossi},
  \citenamefont {Shegai}, \citenamefont {Erhart},\ and\ \citenamefont
  {Antosiewicz}}]{Rossi2019}%
  \BibitemOpen
  \bibfield  {author} {\bibinfo {author} {\bibfnamefont {T.~P.}\ \bibnamefont
  {Rossi}}, \bibinfo {author} {\bibfnamefont {T.}~\bibnamefont {Shegai}},
  \bibinfo {author} {\bibfnamefont {P.}~\bibnamefont {Erhart}},\ and\ \bibinfo
  {author} {\bibfnamefont {T.~J.}\ \bibnamefont {Antosiewicz}},\ }\bibfield
  {title} {\bibinfo {title} {Strong plasmon-molecule coupling at the nanoscale
  revealed by first-principles modeling},\ }\href
  {https://doi.org/10.1038/s41467-019-11315-5} {\bibfield  {journal} {\bibinfo
  {journal} {Nat. Commun.}\ }\textbf {\bibinfo {volume} {10}},\ \bibinfo
  {pages} {1} (\bibinfo {year} {2019})}\BibitemShut {NoStop}%
\bibitem [{\citenamefont {Wu}\ \emph {et~al.}(2021)\citenamefont {Wu},
  \citenamefont {Yan},\ and\ \citenamefont {Lalanne}}]{wu2021}%
  \BibitemOpen
  \bibfield  {author} {\bibinfo {author} {\bibfnamefont {T.}~\bibnamefont
  {Wu}}, \bibinfo {author} {\bibfnamefont {W.}~\bibnamefont {Yan}},\ and\
  \bibinfo {author} {\bibfnamefont {P.}~\bibnamefont {Lalanne}},\ }\bibfield
  {title} {\bibinfo {title} {Bright plasmons with cubic nanometer mode volumes
  through mode hybridization},\ }\href
  {https://doi.org/10.1021/acsphotonics.0c01569} {\bibfield  {journal}
  {\bibinfo  {journal} {ACS Photonics}\ }\textbf {\bibinfo {volume} {8}},\
  \bibinfo {pages} {307} (\bibinfo {year} {2021})}\BibitemShut {NoStop}%
\bibitem [{\citenamefont {Ros\l{}awska}\ \emph {et~al.}(2021)\citenamefont
  {Ros\l{}awska}, \citenamefont {Merino}, \citenamefont {Grewal}, \citenamefont
  {Leon}, \citenamefont {Kuhnke},\ and\ \citenamefont
  {Kern}}]{roslawska2021atomicscale}%
  \BibitemOpen
  \bibfield  {author} {\bibinfo {author} {\bibfnamefont {A.}~\bibnamefont
  {Ros\l{}awska}}, \bibinfo {author} {\bibfnamefont {P.}~\bibnamefont
  {Merino}}, \bibinfo {author} {\bibfnamefont {A.}~\bibnamefont {Grewal}},
  \bibinfo {author} {\bibfnamefont {C.~C.}\ \bibnamefont {Leon}}, \bibinfo
  {author} {\bibfnamefont {K.}~\bibnamefont {Kuhnke}},\ and\ \bibinfo {author}
  {\bibfnamefont {K.}~\bibnamefont {Kern}},\ }\href@noop {} {\bibinfo {title}
  {Atomic-scale structural fluctuations of a plasmonic cavity}} (\bibinfo
  {year} {2021}),\ \Eprint {https://arxiv.org/abs/2102.11671} {arXiv:2102.11671
  [cond-mat.mes-hall]} \BibitemShut {NoStop}%
\bibitem [{\citenamefont {Valiev}\ \emph {et~al.}(2010)\citenamefont {Valiev},
  \citenamefont {Bylaska}, \citenamefont {Govind}, \citenamefont {Kowalski},
  \citenamefont {Straatsma}, \citenamefont {Van~Dam}, \citenamefont {Wang},
  \citenamefont {Nieplocha}, \citenamefont {Apra}, \citenamefont {Windus} \emph
  {et~al.}}]{valiev2010nwchem}%
  \BibitemOpen
  \bibfield  {author} {\bibinfo {author} {\bibfnamefont {M.}~\bibnamefont
  {Valiev}}, \bibinfo {author} {\bibfnamefont {E.~J.}\ \bibnamefont {Bylaska}},
  \bibinfo {author} {\bibfnamefont {N.}~\bibnamefont {Govind}}, \bibinfo
  {author} {\bibfnamefont {K.}~\bibnamefont {Kowalski}}, \bibinfo {author}
  {\bibfnamefont {T.~P.}\ \bibnamefont {Straatsma}}, \bibinfo {author}
  {\bibfnamefont {H.~J.}\ \bibnamefont {Van~Dam}}, \bibinfo {author}
  {\bibfnamefont {D.}~\bibnamefont {Wang}}, \bibinfo {author} {\bibfnamefont
  {J.}~\bibnamefont {Nieplocha}}, \bibinfo {author} {\bibfnamefont
  {E.}~\bibnamefont {Apra}}, \bibinfo {author} {\bibfnamefont {T.~L.}\
  \bibnamefont {Windus}}, \emph {et~al.},\ }\bibfield  {title} {\bibinfo
  {title} {Nwchem: a comprehensive and scalable open-source solution for large
  scale molecular simulations},\ }\href@noop {} {\bibfield  {journal} {\bibinfo
   {journal} {Comput. Phys. Commun.}\ }\textbf {\bibinfo {volume} {181}},\
  \bibinfo {pages} {1477} (\bibinfo {year} {2010})}\BibitemShut {NoStop}%
\bibitem [{\citenamefont {Sibata}\ \emph {et~al.}(2004)\citenamefont {Sibata},
  \citenamefont {Tedesco},\ and\ \citenamefont {Marchetti}}]{SIBATA2004131}%
  \BibitemOpen
  \bibfield  {author} {\bibinfo {author} {\bibfnamefont {M.}~\bibnamefont
  {Sibata}}, \bibinfo {author} {\bibfnamefont {A.}~\bibnamefont {Tedesco}},\
  and\ \bibinfo {author} {\bibfnamefont {J.}~\bibnamefont {Marchetti}},\
  }\bibfield  {title} {\bibinfo {title} {Photophysicals and photochemicals
  studies of zinc(ii) phthalocyanine in long time circulation micelles for
  photodynamic therapy use},\ }\href
  {https://doi.org/https://doi.org/10.1016/j.ejps.2004.06.004} {\bibfield
  {journal} {\bibinfo  {journal} {Eur. J. Pharm. Sci.}\ }\textbf {\bibinfo
  {volume} {23}},\ \bibinfo {pages} {131} (\bibinfo {year} {2004})}\BibitemShut
  {NoStop}%
\bibitem [{\citenamefont {Caplins}\ \emph {et~al.}(2016)\citenamefont
  {Caplins}, \citenamefont {Mullenbach}, \citenamefont {Holmes},\ and\
  \citenamefont {Blank}}]{Caplins2016}%
  \BibitemOpen
  \bibfield  {author} {\bibinfo {author} {\bibfnamefont {B.~W.}\ \bibnamefont
  {Caplins}}, \bibinfo {author} {\bibfnamefont {T.~K.}\ \bibnamefont
  {Mullenbach}}, \bibinfo {author} {\bibfnamefont {R.~J.}\ \bibnamefont
  {Holmes}},\ and\ \bibinfo {author} {\bibfnamefont {D.~A.}\ \bibnamefont
  {Blank}},\ }\bibfield  {title} {\bibinfo {title} {Femtosecond to nanosecond
  excited state dynamics of vapor deposited copper phthalocyanine thin films},\
  }\href {https://doi.org/10.1039/C6CP00958A} {\bibfield  {journal} {\bibinfo
  {journal} {Phys. Chem. Chem. Phys.}\ }\textbf {\bibinfo {volume} {18}},\
  \bibinfo {pages} {11454} (\bibinfo {year} {2016})}\BibitemShut {NoStop}%
\bibitem [{\citenamefont {Delga}\ \emph {et~al.}(2014)\citenamefont {Delga},
  \citenamefont {Feist}, \citenamefont {Bravo-Abad},\ and\ \citenamefont
  {Garcia-Vidal}}]{Delga_2014}%
  \BibitemOpen
  \bibfield  {author} {\bibinfo {author} {\bibfnamefont {A.}~\bibnamefont
  {Delga}}, \bibinfo {author} {\bibfnamefont {J.}~\bibnamefont {Feist}},
  \bibinfo {author} {\bibfnamefont {J.}~\bibnamefont {Bravo-Abad}},\ and\
  \bibinfo {author} {\bibfnamefont {F.~J.}\ \bibnamefont {Garcia-Vidal}},\
  }\bibfield  {title} {\bibinfo {title} {Theory of strong coupling between
  quantum emitters and localized surface plasmons},\ }\href
  {https://doi.org/10.1088/2040-8978/16/11/114018} {\bibfield  {journal}
  {\bibinfo  {journal} {J. Opt.}\ }\textbf {\bibinfo {volume} {16}},\ \bibinfo
  {pages} {114018} (\bibinfo {year} {2014})}\BibitemShut {NoStop}%
\bibitem [{\citenamefont {Neuman}\ \emph {et~al.}(2018)\citenamefont {Neuman},
  \citenamefont {Esteban}, \citenamefont {Casanova}, \citenamefont
  {{Garc{\'i}a-Vidal}},\ and\ \citenamefont {Aizpurua}}]{Neuman2018}%
  \BibitemOpen
  \bibfield  {author} {\bibinfo {author} {\bibfnamefont {T.}~\bibnamefont
  {Neuman}}, \bibinfo {author} {\bibfnamefont {R.}~\bibnamefont {Esteban}},
  \bibinfo {author} {\bibfnamefont {D.}~\bibnamefont {Casanova}}, \bibinfo
  {author} {\bibfnamefont {F.~J.}\ \bibnamefont {{Garc{\'i}a-Vidal}}},\ and\
  \bibinfo {author} {\bibfnamefont {J.}~\bibnamefont {Aizpurua}},\ }\bibfield
  {title} {\bibinfo {title} {Coupling of {{Molecular Emitters}} and {{Plasmonic
  Cavities}} beyond the {{Point}}-{{Dipole Approximation}}},\ }\href
  {https://doi.org/10.1021/acs.nanolett.7b05297} {\bibfield  {journal}
  {\bibinfo  {journal} {Nano Lett.}\ }\textbf {\bibinfo {volume} {18}},\
  \bibinfo {pages} {2358} (\bibinfo {year} {2018})}\BibitemShut {NoStop}%
\bibitem [{\citenamefont {Cao}\ \emph {et~al.}(2021)\citenamefont {Cao},
  \citenamefont {Ros{\l}awska}, \citenamefont {Doppagne}, \citenamefont
  {Romeo}, \citenamefont {F{\'e}ron}, \citenamefont {Ch{\'e}rioux},
  \citenamefont {Bulou}, \citenamefont {Scheurer},\ and\ \citenamefont
  {Schull}}]{Cao2021}%
  \BibitemOpen
  \bibfield  {author} {\bibinfo {author} {\bibfnamefont {S.}~\bibnamefont
  {Cao}}, \bibinfo {author} {\bibfnamefont {A.}~\bibnamefont {Ros{\l}awska}},
  \bibinfo {author} {\bibfnamefont {B.}~\bibnamefont {Doppagne}}, \bibinfo
  {author} {\bibfnamefont {M.}~\bibnamefont {Romeo}}, \bibinfo {author}
  {\bibfnamefont {M.}~\bibnamefont {F{\'e}ron}}, \bibinfo {author}
  {\bibfnamefont {F.}~\bibnamefont {Ch{\'e}rioux}}, \bibinfo {author}
  {\bibfnamefont {H.}~\bibnamefont {Bulou}}, \bibinfo {author} {\bibfnamefont
  {F.}~\bibnamefont {Scheurer}},\ and\ \bibinfo {author} {\bibfnamefont
  {G.}~\bibnamefont {Schull}},\ }\bibfield  {title} {\bibinfo {title} {Energy
  funneling within multi-chromophore architectures monitored with sub-nanometer
  resolution},\ }\bibfield  {journal} {\bibinfo  {journal} {Nat. Chemistry}\
  }\href {https://doi.org/10.1038/s41557-021-00697-z}
  {10.1038/s41557-021-00697-z} (\bibinfo {year} {2021})\BibitemShut {NoStop}%
\bibitem [{\citenamefont {Dutreix}\ \emph {et~al.}(2020)\citenamefont
  {Dutreix}, \citenamefont {Avriller}, \citenamefont {Lounis},\ and\
  \citenamefont {Pistolesi}}]{Dutreix2020}%
  \BibitemOpen
  \bibfield  {author} {\bibinfo {author} {\bibfnamefont {C.}~\bibnamefont
  {Dutreix}}, \bibinfo {author} {\bibfnamefont {R.}~\bibnamefont {Avriller}},
  \bibinfo {author} {\bibfnamefont {B.}~\bibnamefont {Lounis}},\ and\ \bibinfo
  {author} {\bibfnamefont {F.}~\bibnamefont {Pistolesi}},\ }\bibfield  {title}
  {\bibinfo {title} {Two-level system as topological actuator for
  nanomechanical modes},\ }\href
  {https://doi.org/10.1103/PhysRevResearch.2.023268} {\bibfield  {journal}
  {\bibinfo  {journal} {Phys. Rev. Research}\ }\textbf {\bibinfo {volume}
  {2}},\ \bibinfo {pages} {023268} (\bibinfo {year} {2020})}\BibitemShut
  {NoStop}%
\bibitem [{\citenamefont {Das}\ \emph {et~al.}(2017)\citenamefont {Das},
  \citenamefont {Elfving}, \citenamefont {Faez},\ and\ \citenamefont
  {S\o{}rensen}}]{Sumanta2017}%
  \BibitemOpen
  \bibfield  {author} {\bibinfo {author} {\bibfnamefont {S.}~\bibnamefont
  {Das}}, \bibinfo {author} {\bibfnamefont {V.~E.}\ \bibnamefont {Elfving}},
  \bibinfo {author} {\bibfnamefont {S.}~\bibnamefont {Faez}},\ and\ \bibinfo
  {author} {\bibfnamefont {A.~S.}\ \bibnamefont {S\o{}rensen}},\ }\bibfield
  {title} {\bibinfo {title} {Interfacing superconducting qubits and single
  optical photons using molecules in waveguides},\ }\href
  {https://doi.org/10.1103/PhysRevLett.118.140501} {\bibfield  {journal}
  {\bibinfo  {journal} {Phys. Rev. Lett.}\ }\textbf {\bibinfo {volume} {118}},\
  \bibinfo {pages} {140501} (\bibinfo {year} {2017})}\BibitemShut {NoStop}%
\bibitem [{\citenamefont {Laucht}\ \emph {et~al.}(2009)\citenamefont {Laucht},
  \citenamefont {Hofbauer}, \citenamefont {Hauke}, \citenamefont {Angele},
  \citenamefont {Stobbe}, \citenamefont {Kaniber}, \citenamefont {B\"{o}hm},
  \citenamefont {Lodahl}, \citenamefont {Amann},\ and\ \citenamefont
  {Finley}}]{Laucht2009}%
  \BibitemOpen
  \bibfield  {author} {\bibinfo {author} {\bibfnamefont {A.}~\bibnamefont
  {Laucht}}, \bibinfo {author} {\bibfnamefont {F.}~\bibnamefont {Hofbauer}},
  \bibinfo {author} {\bibfnamefont {N.}~\bibnamefont {Hauke}}, \bibinfo
  {author} {\bibfnamefont {J.}~\bibnamefont {Angele}}, \bibinfo {author}
  {\bibfnamefont {S.}~\bibnamefont {Stobbe}}, \bibinfo {author} {\bibfnamefont
  {M.}~\bibnamefont {Kaniber}}, \bibinfo {author} {\bibfnamefont
  {G.}~\bibnamefont {B\"{o}hm}}, \bibinfo {author} {\bibfnamefont
  {P.}~\bibnamefont {Lodahl}}, \bibinfo {author} {\bibfnamefont {M.-C.}\
  \bibnamefont {Amann}},\ and\ \bibinfo {author} {\bibfnamefont {J.~J.}\
  \bibnamefont {Finley}},\ }\bibfield  {title} {\bibinfo {title} {Electrical
  control of spontaneous emission and strong coupling for a single quantum
  dot},\ }\href {https://doi.org/10.1088/1367-2630/11/2/023034} {\bibfield
  {journal} {\bibinfo  {journal} {New Journal of Physics}\ }\textbf {\bibinfo
  {volume} {11}},\ \bibinfo {pages} {023034} (\bibinfo {year}
  {2009})}\BibitemShut {NoStop}%
\bibitem [{\citenamefont {Breuer}\ and\ \citenamefont
  {Petruccione}(2003)}]{Breuer2003}%
  \BibitemOpen
  \bibfield  {author} {\bibinfo {author} {\bibfnamefont {H.-P.}\ \bibnamefont
  {Breuer}}\ and\ \bibinfo {author} {\bibfnamefont {F.}~\bibnamefont
  {Petruccione}},\ }\href@noop {} {\emph {\bibinfo {title} {The theory of open
  quantum systems}}}\ (\bibinfo  {publisher} {Oxford university press},\
  \bibinfo {year} {2003})\BibitemShut {NoStop}%
\bibitem [{\citenamefont {Johnson}\ and\ \citenamefont
  {Christy}(1972)}]{JohnsonChristy1972}%
  \BibitemOpen
  \bibfield  {author} {\bibinfo {author} {\bibfnamefont {P.~B.}\ \bibnamefont
  {Johnson}}\ and\ \bibinfo {author} {\bibfnamefont {R.~W.}\ \bibnamefont
  {Christy}},\ }\bibfield  {title} {\bibinfo {title} {Optical constants of the
  noble metals},\ }\href {https://doi.org/10.1103/PhysRevB.6.4370} {\bibfield
  {journal} {\bibinfo  {journal} {Phys. Rev. B}\ }\textbf {\bibinfo {volume}
  {6}},\ \bibinfo {pages} {4370} (\bibinfo {year} {1972})}\BibitemShut
  {NoStop}%
\end{thebibliography}%

\end{document}